# A low-cost four-component relativistic equation of motion coupled cluster method based on frozen natural spinors: Theory, Implementation and Benchmark

Kshitijkumar Surjuse[1], Somesh Chamoli[1], Malaya K. Nayak[2,3,a] and Achintya Kumar Dutta[1,b]

[1]Department of Chemistry, Indian Institute of Technology Bombay, Powai, Mumbai 400076, India

2)Theoretical Chemistry Section, Bhabha Atomic Research Centre, Trombay, Mumbai 400085, India

3)Homi Bhabha National Institute, BARC Training School Complex, Anushakti Nagar, Mumbai 400094, India

We present the theory and the implementation of a low-cost four-component relativistic equation of motion coupled cluster method for ionized states based on frozen natural spinors. A single threshold (natural spinor occupancy) can control the accuracy of the calculated ionization potential values. Frozen natural spinors can significantly reduce the computational cost for valence and core-ionization energies with systematically controllable accuracy. The convergence of the ionization potential values with respect to the natural spinor occupancy threshold becomes slower with the increase in basis set dimension. However, the use of a natural spinor threshold of $10^{-5}$ and $10^{-6}$ gives excellent agreement with experimental results for valence and core ionization energies, respectively.

a)Electronic mail: mknayak@barc.gov.in; mk.nayak72@gmail.com

b)Electronic mail: achintya@chem.iitb.ac.in

# 1. Introduction

The ionization of atoms and molecules can lead to interesting properties, which has fascinated the generation of physicists and chemists. Theoretical simulations can be of significant help in the interpretation of the experimentally measured ionization-induced phenomenon. Accurate simulation of atomic and molecular spectroscopy requires a balanced inclusion of relativistic and correlation effects, especially for systems containing heavy elements. The four-component Dirac-Hartree-Fock (DHF) method provides an efficient framework to include relativistic effects in quantum chemical calculations. However, the single determinant picture of the DHF is not often enough to give quantitative accuracy, and one needs to go for electron correlation methods. Among the various relativistic electron correlation methods available, the coupled-cluster method is considered to be one of the most accurate and systematically improvable[1]. The coupled cluster method is generally used in singles and doubles approximation (CCSD), and it scales the $O(N^6)$ power of the basis set.

The calculation of the difference of energies like ionization potential (IP) and electron affinity (EA) within the coupled cluster framework can be performed in two ways. The first approach is the so-called Δ-based method, where two separate calculations are performed on the neutral and cationic/anionic systems. Subsequently, the IP/EA values are calculated as the difference between the two energies. Alternatively, one can calculate the IP/EA in a single calculation using the equation of motion approach[2]. The equation of motion coupled cluster theory (EOM-CC) gives identical results[3–5] to the coupled cluster linear response approach[6] and allows one to explore various sectors of Fock-space without going into the theoretical intricacies of Fock-space multi-reference coupled cluster theory[7–10]. The EOM-CC method based on four component Dirac-Coulomb Hamiltonian[11,12] has shown to give excellent agreement with available experimental results, even with singles and doubles truncation (EOM-CCSD) of cluster operator. However, the high computational cost of coupled cluster method restricts its uses beyond small systems and/or moderate basis sets. The problem is aggravated in the relativistic case, where the presence of complex integrals further increases the computational cost. A four-component relativistic CCSD calculation on a closed shell atom or molecule will be at least 32 times[13] more costly than the corresponding non-relativistic version in the same basis set. Therefore, it is essential to reduce the computational cost of the relativistic IP-EOM-CCSD method to increase its domain of applicability. Massively parallel implementation of the relativistic coupled cluster, which can scale on multiple CPU and GPU, has also been reported[14]. Attempts have been made to reduce the intrinsic computational cost of relativistic EOM-CCSD methods by perturbational truncation[11,15] of the cluster amplitudes. However, the recently developed frozen natural spinor framework gives a more systematic approach to reduce the computation cost of coupled cluster calculations with systematically controllable accuracy[16]. It has been shown that a single threshold can control the accuracy of the FNS-CCSD method for both energy and properties of close shell molecules. Visscher and co-workers reported a similar development within the framework of the two-component relativistic Hamiltonian[17]. The aim of this paper is to extend the frozen natural spinor framework to the four-component relativistic IP-EOM-CCSD method.

## 2. Theory
### 2.1 Relativistic Coupled Cluster Method

The exact wave function in the relativistic coupled cluster method is generated as

$$\psi_{cc} = e^{\hat{T}} |\phi_0\rangle \tag{1}$$

with, $\phi_0$ as the N-electron single determinant Dirac-Hartree-Fock (DHF) wave function and $\hat{T}$ is the coupled-cluster operator given by

$$\hat{T} = \hat{T}_1 + \hat{T}_2 \ldots + \hat{T}_N \tag{2}$$

Where, any general n-spinor cluster operator $\hat{T}_n$ is expressed in the second quantized notation as

$$\hat{T}_n = \left(\frac{1}{n!}\right)^2 \sum_{ijk\ldots abc\ldots} t_{ijk\ldots}^{abc\ldots} \{\hat{a}_a^\dagger \hat{a}_b^\dagger \hat{a}_c^\dagger \ldots \hat{a}_i \hat{a}_j \hat{a}_k \ldots\} \tag{3}$$

Here, (i, j, k, l), (a, b, c, d), and (p, q, r, s) are used to denote occupied, virtual, and any general spinors, respectively. The $t_{ijk\ldots}^{abc\ldots}$ are coupled cluster amplitudes and $\hat{a}$ and $\hat{a}^\dagger$ are hole and particle creation operators with respect to the DHF vacuum.

Truncating the $\hat{T}$ operator to singles and doubles excitation level yields the coupled cluster singles and doubles (CCSD) method.

$$\hat{T} = \hat{T}_1 + \hat{T}_2 \tag{4}$$

$$\hat{T}_1 = \sum_{ia} t_i^a \{\hat{a}_a^\dagger \hat{a}_i\} \tag{5}$$

$$\hat{T}_2 = \frac{1}{2} \sum_{ijab} t_{ij}^{ab} \{\hat{a}_a^\dagger \hat{a}_b^\dagger \hat{a}_i \hat{a}_j\} \tag{6}$$

The coupled cluster amplitudes in CCSD method are obtained by solving the following simultaneous non-linear equations

$$\langle \phi_i^a | e^{-\hat{T}} \hat{H}_{DC} e^{\hat{T}} | \phi_0 \rangle = 0 \tag{7}$$

$$\langle \phi_{ij}^{ab} | e^{-\hat{T}} \hat{H}_{DC} e^{\hat{T}} | \phi_0 \rangle = 0 \tag{8}$$

With $\phi_i^a$ and $\phi_{ij}^{ab}$ are singly and doubly excited determinants and $\hat{H}_{DC}$ is the Dirac-Coulomb Hamiltonian.

$$\hat{H}_{DC} = \sum_i^N \left[ c\alpha_i \cdot p_i + \beta_i m_0 c^2 + \sum_A^{N_{nuc}} V_{iA} \right] + \sum_{i>j}^N \frac{1}{r_{ij}} I_4 \tag{9}$$

Where $\alpha$ and $\beta$ are Dirac matrices, $V_{iA}$ is the potential energy operator for the $i^{th}$ electron in the field of nucleus A, $m_0$ is the rest mass of a free electron and $c$ is the speed of light. $I_4$ is 4×4 identity matrix. The summation in equation (9) only encompasses positive energy electrons with the proper consideration of no-pair approximation.

One can augment the DC Hamiltonian with the Gaunt term for electromagnetic interactions and with the Breit term for electron retardation effects in addition to electromagnetic interactions.

$$\hat{H}_{DCG} = \sum_i^N \left[ c\alpha_i \cdot p_i + \beta_i m_0 c^2 + \sum_A^{N_{nuc}} V_{iA} \right] + \sum_{i>j}^N \left( \frac{1}{r_{ij}} + G_{ij} \right) I_4 \tag{10}$$

and

$$\hat{H}_{DCB} = \sum_i^N \left[ c\alpha_i \cdot p_i + \beta_i m_0 c^2 + \sum_A^{N_{nuc}} V_{iA} \right] + \sum_{i>j}^N \left( \frac{1}{r_{ij}} + B_{ij} \right) I_4 \tag{11}$$

Where,

$$G_{ij} = -\frac{\alpha_i \cdot \alpha_j}{2r_{ij}} \tag{12}$$

and

$$B_{ij} = -\frac{1}{2r_{ij}} \left[ \alpha_i \cdot \alpha_j + \frac{(\alpha_i \times r_{ij}) \cdot (\alpha_j \times r_{ij})}{r_{ij}^2} \right] \tag{13}$$

Also, the non-Hermitian similarity transformed Hamiltonian $\bar{H}$ can be introduced as

$$\bar{H} = e^{-\hat{T}} \hat{H}_{DC} e^{\hat{T}} \tag{14}$$

And the expression for the energy is:

$$E_{CCSD} = \langle \phi_0 | \bar{H} | \phi_0 \rangle \tag{15}$$

## 2.2 Equation of motion coupled cluster (EOM-CC)

In the equation of motion coupled cluster method, the target states are obtained by the diagonalization of the non-Hermitian similarity transformed Hamiltonian $\bar{H}$ within the space of suitably chosen configurations. The non-hermiticity of $\bar{H}$ leads to left $\langle \phi_0 | \hat{L}$ and right $\hat{R} | \phi_0 \rangle$ eigenvectors which are not adjoint of each other. i.e.,

$$\langle \phi_0 | \hat{L} \neq \left( \hat{R} | \phi_0 \rangle \right)^\dagger \tag{16}$$

These two sets of left and right eigenvectors are biorthogonal

$$\langle \phi_0 | \hat{L}_i \hat{R}_j | \phi_0 \rangle = \delta_{ij} \tag{17}$$

and can be obtained by solving two different eigenvalue equations.

$$\bar{H} \hat{R}_k | \phi_0 \rangle = E_k \hat{R}_k | \phi_0 \rangle \tag{18}$$

$$\langle \phi_0 | \hat{L}_k \bar{H} = \langle \phi_0 | \hat{L}_k E_k \quad (19)$$

It is enough to solve either left or right eigenvector equation if we are only interested in the target state energies. However, one needs to solve both for property calculations. The exact form of the $\hat{L}_k$ and $\hat{R}_k$ depends upon the nature of the target state. For the ionization energy problem, they have the following form

$$\hat{R}^{IP} = \sum_i r_i \{\hat{a}_i\} + \sum_{i>j,a} r_{ij}^a \{\hat{a}_a^\dagger \hat{a}_j \hat{a}_i\} + \ldots \quad (20)$$

$$\hat{L}^{IP} = \sum_i l^i \{\hat{a}_i^\dagger\} + \sum_{i>j,a} l_a^{ij} \{\hat{a}_j^\dagger \hat{a}_i^\dagger \hat{a}_a\} + \ldots \quad (21)$$

The $\hat{L}$ and $\hat{R}$ operators are generally used in singles and doubles truncation (IP-EOM-CCSD) schemes. The difference in the energy between the target and the reference state (i.e. $\omega_k = E_k - E_0$) can be obtained directly by using the commutator form of eq 18.

$$[\bar{H}, \hat{R}_k] | \phi_0 \rangle = \omega_k \hat{R}_k | \phi_0 \rangle \quad (22)$$

The equation (22) is generally solved using Davidson's iterative Diagonalization method[18]. The construction of the sigma vectors in the Davidson's iterative Diagonalization method for the IP-EOM-CCSD scales as $O(N_O^3 N_V^2)$ power of the basis set for each root. Where the $N_O$ is the no of occupied spinors and $N_V$ is the no of virtual spinors. The ground CCSD calculation scales as $O(N_O^2 N_V^4)$ and is the computationally most expensive part in a IP-EOM-CCSD calculations.

## 2.3 Frozen Natural Spinors

The natural spinors are the eigenfunctions of a spin-coupled correlated one-body reduced density matrix obtained from relativistic correlated wavefunction calculations. They are the relativistic analog of the natural orbitals proposed by Löwdin[19]. The various flavors of natural orbitals, starting from frozen natural orbitals[20–25], natural transition orbitals[26–28], and pair-natural orbitals[29–33] have been used to reduce the virtual space size in the non-relativistic wave function based method. Landau *et al.* have reported[22] the frozen natural orbital-based implementation of non-relativistic IP-EOM-CCSD, where the natural orbitals are generated from the MP2 calculations. Similarly, within the four-component relativistic methods, the MP2 method in the no-pair approximation is used for generating natural spinors[16] for its favorable scaling. The virtual-virtual block of the one-body reduced density matrix is constructed as

$$D_{ab} = \sum_{ijc} \frac{\langle ac || ij \rangle \langle ij || bc \rangle}{\varepsilon_{ij}^{ac} \varepsilon_{ij}^{bc}} \quad (23)$$

where

$$\varepsilon_{ij}^{ac} = \varepsilon_i + \varepsilon_j - \varepsilon_a - \varepsilon_c \qquad (24)$$

$$\varepsilon_{ij}^{bc} = \varepsilon_i + \varepsilon_j - \varepsilon_b - \varepsilon_c \qquad (25)$$

and $\varepsilon_i, \varepsilon_j, \varepsilon_a, \varepsilon_b$ and $\varepsilon_c$ are molecular spinor energies, and $\langle ac||ij \rangle$ and $\langle ij||bc \rangle$ are the antisymmetrized two-electron integrals obtained from DHF method. The virtual natural spinors $V$ and their corresponding occupation numbers are obtained from the diagonalization of the virtual-virtual block of the one-body reduced density matrix.

$$D_{ab}V = Vn \qquad (26)$$

The occupation number of the virtual natural spinor conveys its importance towards the correlation energy. Hence these virtual natural spinors are sorted in descending order of their occupation numbers, and the virtual space can be reduced by truncating the virtual natural spinors with occupation numbers below a certain threshold. The virtual-virtual block of the Fock Matrix is then transformed into the truncated natural spinor basis.

$$F_{vv}^{NS} = \overline{V}^{\dagger} F_{vv} \overline{V} \qquad (27)$$

Where $F_{vv}^{NS}$ represents the virtual-virtual block of the Fock matrix in the truncated virtual natural spinor basis, $\overline{V}$ is the truncated virtual natural spinors, and the $F_{vv}$ denotes the virtual-virtual block of the initial Fock matrix. Diagonalization of $F_{vv}^{NS}$ gives the semi-canonical virtual natural spinors and their corresponding orbital energies.

$$F_{vv}^{NS} \overline{Z} = \overline{Z} \varepsilon_{NS} \qquad (28)$$

$\overline{Z}$ represents the semi-canonical virtual natural spinors, and $\varepsilon_{NS}$ are their orbital energies. The transformation from canonical DHF virtual spinor space to semi-canonical virtual natural space can be expressed as the product of $\overline{Z}$ and $\overline{V}$

$$U = \overline{Z}\overline{V} \qquad (29)$$

Hence, in our basis, occupied spinors remain the same as canonical DHF spinors, and the virtual spinors are transformed into semi-canonical virtual natural spinors. This approach is denoted as frozen natural spinors (FNS) following its non-relativistic analog. The subsequent coupled cluster and IP-EOM calculations are performed in the frozen natural spinor basis. The ground state calculation in the FNS-IP-EOM-CCSD method scales as $O(N_O^2(N_V - N_{FV})^4)$ and the EOM part scales as $O(N_O^2(N_V - N_{FV})^3)$ power of the basis set.

## 2.4 Core Valence Separation (CVS)

The states formed by the single ionization from core orbitals remain embedded in the continuum of the satellite states formed due to the ionization from valence orbitals. This leads to convergence issues in the Davidson iterative diagonalization procedure. One of the easy solutions to the problem is to use the core-valence separation approach of Domcke and Cederbaum[34]. In CVS approximation, the core and valence electrons are decoupled based on their difference in energy. Coriani and Koch have extended the CVS approximation to coupled cluster based excited-state methods[35]. Gomes and co-worker have recently reported a CVS-EOM-CCSD method for core excited and core-ionized states based on a four-component Dirac-Coulomb (-Gaunt) Hamiltonian[36]. In the present implementation, the core-valence separation is achieved by applying a projection operator P on each iteration of the Davidson procedure during the EOM-CC solution, which selectively zeros out the valence contributions to the EOM sigma vectors. Which results in

$$\sigma_i = \sigma_{ij}^a = 0 \text{ if } i, j \in valence \tag{30}$$

## 2.5 Implementation and Computational Details

The FNS-IP-EOM-CCSD has been implemented in the development version of BAGH[37]. The converged four-component DHF spinors[38], along with the one and two-electron integrals, are taken from the PYSCF[39]. In the first step, the two-electron molecular integrals with two external indices required for the MP2 calculations are generated in the canonical spinor basis using a partial integral transformation.

$$(ia|jb) = \sum_{\mu\nu\kappa\lambda} C_{i\mu}^{L*} C_{a\nu}^{L} C_{j\kappa}^{L*} C_{b\lambda}^{L} \left(\mu^L \nu^L | \kappa^L \lambda^L\right) + \sum_{\mu\nu\kappa\lambda} C_{i\mu}^{S*} C_{a\nu}^{S} C_{j\kappa}^{S*} C_{b\lambda}^{S} \left(\mu^S \nu^S | \kappa^S \lambda^S\right)$$
$$+ \sum_{\mu\nu\kappa\lambda} C_{i\mu}^{L*} C_{a\nu}^{L} C_{j\kappa}^{S*} C_{b\lambda}^{S} \left(\mu^L \nu^L | \kappa^S \lambda^S\right) + \sum_{\mu\nu\kappa\lambda} C_{i\mu}^{S*} C_{a\nu}^{S} C_{j\kappa}^{L*} C_{b\lambda}^{L} \left(\mu^S \nu^S | \kappa^L \lambda^L\right) \tag{31}$$

where i, j, a, and b are the molecular spinors; κ, λ, μ, and ν are the atomic spinors; L and S denote the large and small component basis set; $C$ denotes the atomic spinor to molecular spinor transformation coefficients.

The antisymmetrization and reordering are done after the transformation is completed.

$$\langle ij \| ab \rangle = (ia|jb) - (ib|ja) \tag{32}$$

In the next step, the MP2 natural spinors are calculated following equations (23-29). The atomic spinor to natural spinor transformation matrix is generated as

$$C'^{L}_{p'\mu} = U_{p'x} C^{L}_{x\mu} \tag{33}$$

$$C'^{S}_{p'\mu} = U_{p'x} C^{S}_{x\mu} \tag{34}$$

Where $x$ denotes molecular spinor and $p'$ denotes frozen natural spinor. The $C'$ denotes the transformation matrix from atomic spinor to frozen natural spinor. Subsequently, 0-4 particle(external) integrals are generated in frozen natural spinor basis and antisymmetrized.

$$(p'r'|q's') = \sum_{\mu\nu\kappa\lambda} C'^{L*}_{p'\mu} C'^{L}_{r'\nu} C'^{L*}_{q'\kappa} C'^{L}_{s\lambda} \left(\mu^L \nu^L | \kappa^L \lambda^L\right) + \sum_{\mu\nu\kappa\lambda} C'^{S*}_{p'\mu} C'^{S}_{r'\nu} C'^{S*}_{q'\kappa} C'^{S}_{s\lambda} \left(\mu^S \nu^S | \kappa^S \lambda^S\right)$$
$$+ \sum_{\mu\nu\kappa\lambda} C'^{L*}_{p'\mu} C'^{L}_{r'\nu} C'^{S*}_{q'\kappa} C'^{S}_{s\lambda} \left(\mu^L \nu^L | \kappa^S \lambda^S\right) + \sum_{\mu\nu\kappa\lambda} C'^{S*}_{p'\mu} C'^{S}_{r'\nu} C'^{L*}_{q'\kappa} C'^{L}_{s\lambda} \left(\mu^S \nu^S | \kappa^L \lambda^L\right) \tag{35}$$

$$\langle pq \| rs \rangle = (pr|qs) - (pr|sq) \tag{36}$$

The contribution of Gaunt and Breit term is not included in the integral transformation step and is only considered in the DHF step. In the original implementation of FNS-CCSD[16], we have used the DIRAC[40] interface of BAGH to generate converged DHF spinors and the 2-electron molecular spinor integrals. The complete set of 0-4 particle integral was generated in the canonical spinor basis and subsequently transformed to the FNS basis. The present implementation only requires the generation of 2-external integrals in the canonical basis, and the rest of the integrals are generated directly in the frozen natural spinor basis. It leads to considerable speed up in the integral transformation step and drastically reduces the storage requirement. Figure 1 gives a schematic description of the algorithm for the FNS-IP-EOM-CCSD method.

To investigate the performance of the FNS-IP-EOM-CCSD method for valence IP calculation, we have used the hydrogen halide series (H-X, X=F, Cl, Br, I, and At). Experimental geometry has been used for all the molecules. We have used the aug-cc-pVXZ (X=D, T, Q, and 5) basis set for both hydrogen and fluorine for calculations on HF For HCl aug-cc-pVXZ (X=D, T, and Q) basis set was used for both hydrogen and chlorine. For HBr, HI and HAt calculations were performed using aug-cc-pVXZ (X=D, T, and Q) basis set for hydrogen and dyall.aexz (x=2, 3, and 4) basis set for bromine, iodine, and Astatine. We have used the rare gas series (Ne, Ar, Kr, Xe, and Rn) for the calculations of core ionization potential. The aug-cc-pCVXZ (X=D, T, and Q) basis set for Ne and Ar were used and dyall.aexz (x=2, 3, and 4) basis set were used for Kr, Xe, and Rn. The basis sets were kept uncontracted for all the calculations.

## 3. Results and Discussion

### 3.1 Convergence with respect to the threshold.

Figure 2 gives the convergence of the core IP values with respect to the % of active virtual in FNS and the canonical basis for HF It can be seen that the IP values converge more quickly with respect to the size of the virtual space in the FNS basis than that observed in the canonical virtual spinors. In the case of hydrogen fluoride, the first IP value converges with ~ 60% of the virtual space in the FNS basis. On the hand, to reach the same level of accuracy, one needs to use at least 90% of the virtual space in the canonical basis. The convergence of the IP value with respect to the size of the virtual space seems to be more smooth in the FNS basis than that observed in the canonical basis. One of the most attractive features of the FNS framework[16] is the single criteria of occupation threshold controls the accuracy. In the original implementation paper of FNS based coupled cluster method[16], retention of the virtual natural spinors up to the occupation of $10^{-6}$ was found to be sufficient for getting converged results for ground state energy and property. To find out the optimal FNS threshold for the IP value, we have calculated the convergence of the first valence IP value of HF with respect to the FNS threshold (see Figure 3). The uncontracted aug-cc-pVTZ basis set has been used for the calculation. It can be seen that the valence IP values start to converge on reaching an FNS threshold of $10^{-5}$.

To understand the basis set dependence of the threshold, we have plotted the error in the first IP value of HF with respect to the FNS threshold in the uncontracted aug-cc-pVXZ (X=D, T, Q) basis set (See Figure 4). The plot of the absolute IP value with respect to the FNS threshold is presented in Figure S1. It can be seen that the convergence of error with respect to the FNS threshold is slower in the larger basis sets. Figure 5 presents the percentage of the virtual space kept vs the FNS threshold in the different basis sets. The corresponding values are provided in Table S1. It can be seen that the same threshold generally leads to a smaller percentage of active space in larger basis sets. This leads to slower convergence of IP value with respect to the FNS threshold with an increase in the basis set size. However, an FNS threshold of $10^{-5}$ seems to give reasonable accuracy in all three basis sets.

To check the reproducibility of the trends for heavier elements, we have plotted the convergence of the error in valence IP values with respect to percentage virtual space (See Figure S2) and FNS threshold (Figure S3) for HBr. The uncontracted aug-cc-pVTZ basis set for hydrogen and dyall.ae3z basis set for bromine has been used for calculations. The convergence of the error with respect to the size of the virtual space is faster for HBr than that observed for HF in both FNS and canonical basis. The error in the first valence IP values seems to converge with only 40% of the virtual space in the FNS basis, where one needs to include almost 75% of the virtual space to get similar accuracy in the canonical basis. The convergence of the error in the valence IP value with respect to the FNS threshold is also faster for HBr than that observed for HF The IP values seem to reach convergence with an FNS threshold of $10^{-4}$. For the rest of the calculations on valence IP in the manuscript, we have used an FNS threshold of $10^{-5}$.

In the FNS based implementation of ground state coupled cluster[16], the convergence of the ground state correlation energy with respect to the FNS truncation threshold has been found to be faster in the frozen core approximation than that observed in the all-electron calculations. To understand the effect of frozen core approximation on the convergence of valence IP values, we have plotted the convergence of the error in IP value with respect to the FNS threshold in both frozen core and all-electron calculations of HBr (See Figure 6). The corresponding canonical IP results have been used as the reference value. The convergence of the error with respect to the percentage of virtual space for the same calculation is presented in Figure S4. The difference in the convergence behavior of all-electron and frozen core calculations is not significant for the valence IP This is presumably due to the error calculation between the reference and target state in FNS-IP-EOM-CCSD calculations.

The core-ionized states have considerably different physics than that observed for valence ionized states, and the former is generally associated with a large orbital relaxation effect. In the IP-EOM-CCSD method, the $\hat{R}_2$ operator brings in the orbital relaxation, and one needs to estimate the error in core-ionization energy caused by truncation of the $\hat{R}_2$ in the FNS basis. Figure 7 presents the convergence of the error in K-edge core IP values of HF in the CVS-IP-EOM-CCSD method with respect to the percentage of virtual spinors in canonical and FNS basis. The uncontracted aug-cc-pVTZ basis set has been used for the calculations. It can be seen that the error in core-IP values converges more quickly with respect to the size of the virtual spinors in FNS basis than that observed in the canonical basis. It can be seen that the core IP values converge with only 40% of the virtual space in the FNS basis. On the other hand, to reach a similar level of accuracy, one needs more than 90% of the virtual space in the canonical basis.

Figure 8 presents the convergence of the error in K-edge core IP values of HF with respect to the FNS threshold. It can be seen that the core IP values are overestimated in CVS-IP-EOM-CCSD at very small FNS thresholds and underestimated at the higher threshold. It is contrary to that observed in the case of valence IP values, where the results are systematically underestimated with respect to the canonical values. Figure 9 presents the convergence of error in K-edge core IP values of HF in uncontracted aug-cc-pVXZ (X=D, T and Q) basis set. The convergence of the core IP values with respect to the FNS threshold becomes slower with an increase in the basis set dimension. However, the results seem to converge in all the three basis sets, on reaching a threshold of $10^{-6}$(see Figure S5). To investigate the generality of the trends in the convergence of core IP values with respect to the FNS threshold in heavier elements, we have calculated the K-edge core IP values of Kr. Figure S6 presents the Error in K-edge core IP values in the CVS-IP-EOM-CCSD method with respect to the % of active virtual in canonical and FNS basis for Kr. Uncontracted dyall.ae3z has been used, and canonical CVS-IP-EOM-CCSD results have been used as the reference. The convergence of the core IP values with respect to the size of the virtual space is much faster in the FNS basis than that in the canonical molecular spinor basis. However, one needs to include more than 50 percent of the virtual space to get convergence in the 1s core IP value for Ar, which is much larger than that observed for HF A plot of the error with respect to the FNS threshold (See Figure S7) shows that the results seem to converge with a truncation threshold of $10^{-6}$, the same as the HF It demonstrates that the FNS based CVS-IP-EOM-CCSD methods are near the black box for core-ionization energies and can automatically choose an appropriate virtual space in different atoms and molecules.

### 3.2 Comparision with experiments

To compare the performance of the CVS-IP-EOM-CCSD with experiments, we have chosen the hydrogen halide series (H-X, X=F, Cl, Br, I, and At). Table S2 shows the basis set convergence of HF using uncontracted aug-cc-pVXZ (X=D to 5) basis set. It can be seen that the valence IP values seem to converge on reaching the aug-cc-pVQZ basis set, and the change in the result from aug-cc-pVQZ to aug-cc-pV5Z is negligible. Therefore, the rest of the calculations on valence IP are performed in the QZ. level basis set.

Table 1 presents the valence IP values of the hydrogen halide series. Uncontracted aug-cc-pVXZ(X=D, T, and Q) basis set has been used for H, F, and Cl. Uncontracted dyall.aexz (x=2, 3, and 4) has been used for Br, I, and At. The frozen core approximation has been used for the HI and HAt. The effect of the frozen core approximation on the valence IP values has been discussed in the supporting information. The convergence of the IP values with respect to the basis set size for all the higher halogen halides has been found to be similar to the HF The outer valence IP values in the QZ. level basis set show excellent agreement with the experiment, and the maximum error observed is less than 0.4%. However, the errors for inner valence IP values are much larger than observed for outer valence. The 8 σ state of HBr shows the highest error of 1% with respect to the experimental values. In general, the IP values in the FNS-IP-EOM-CCSD method for inner valence electrons are systematically overestimated with respect to the experimental values. The effect of orbital relaxation on inner valence electrons can be significant. It is well known that the linear excitation operator for the EOM-CC cannot adequately include the orbital relaxation effect[41] in the singles and doubles approximation. One needs to include the effect of the three-body excitation operator in the EOM-CC calculations to have a proper balance between the orbital relaxation and electron correlation effect in IP values. However, the inclusion of triples correction to the FNS based IP-EOM-CCSD method is outside the scope of the present study and will be followed in a separate manuscript. It is important to notice that changes in the valence IP values are significant on going from DZ to TZ quality basis set. The splitting between the two ionized states is almost independent of the basis set used. The effect of the Gaunt and Breit correction on the valence IP values has been found to be small even for H-At (See Table S3).

To understand the basis set convergence behavior of the core-ionization energies, we have calculated K-edge core-ionization energies noble gas series Ne, Ar, Kr, Xe. The uncontracted aug-

cc-pCVXZ(X=D, T, Q) has been used for Ne, Ar, and the uncontracted dyall.aexz (x=2, 3, and 4) basis set has been used for Kr, and Xe. Following the observation of Gomes and co-workers on the importance of (SS|SS) type integrals on core IP values[36], they are calculated analytically for all the calculations in this paper. Table 2 shows that the change in the core-ionization energy with respect to the basis set follows almost the same trend as that of the valence ionization. The change from the TZ to QZ. level is around 0.1 eV for all the molecules, leading to a 0.01% error even for HF.

One can see that the use of a four-component Dirac Coulomb Hamiltonian cannot give sufficient accuracy for core-ionization energy, especially for heavy elements. Xe shows an 0.38% error with respect to the experiment in dyall.ae3z basis set. Gomes and co-workers have shown that one needs to include the higher-order relativistic effects to get a proper agreement with respect to experimental results[36]. Table 3 presents the effect of the inclusion of higher-order relativistic correction (such as Gaunt and Breit) to the core IP values. The uncontracted aug-cc-pCVTZ basis set has been used for Ne, Ar, and dyall.ae3z basis set has been used for Kr, Xe, and Rn. The DC-CVS-FNO-IP-EOM-CCSD method shows the highest error (0.56%) with respect to the experiment for Rn, which is the heaviest element in the noble gas series considered.

The agreement with experiments improves on adding Gaunt correction. The addition of Gaunt correction leads to the redshift for the core IP values for all the elements. The magnitude of the redshift is the smallest for Ne and becomes progressively higher for the heavier elements. The addition of Breit correction led to a Blueshift with respect to the Gaunt corrected value. However, the magnitude of the blue shift is much smaller than the redshift caused by the Gaunt term. Gomes and co-workers have shown that the addition of the QED correction can further improve the accuracy of the core IP values of heavier elements[36]. Therefore, we have added the QED correction from the work of Kaziol and Aucar[42] to our DCB-CVS-FNS-IP-EOM-CCSD results to Kr, Xe, and Rn. The addition of QED correction causes a further redshift of the results, and the error gets reduced to less than 0.1% for all the molecules.

### 3.3 Comparison with non-relativistic results.

One needs to compare the performance of the FNS-IP-EOM-CCSD with the corresponding non-relativistic version. Figure 10 presents the comparison of the error in valence IP value in FNS and FNO-based IP-EOM-CCSD method with respect to the percentage of virtual space for HBr. An uncontracted aug-cc-pVTZ basis set has been used for hydrogen and uncontracted dyall.ae3z basis set has been used for Br. It can be seen that the convergence of valence IP values with respect to the size of the virtual space shows almost identical trends for the four-component relativistic and non-relativistic IP-EOM-CCSD method. Figure 11 presents the convergence of the error in 1s core IP value in FNS and FNO-based IP-EOM-CCSD method with respect to the percentage of virtual space for Kr. Uncontracted dyall.ae3z used for Kr. The trends in the convergence of core IP values with respect to the size of the virtual space are almost identical in the four-component relativistic and non-relativistic IP-EOM-CCSD method, similar to the valence IP.

Table 4 presents the comparison of results in the FNS(FNO)-IP-EOM-CCSD method based on four-component DC Hamiltonian, spin-free X2C, and non-relativistic Hamiltonian for valence IP of HX (X = F, Cl, Br, I, and At). The uncontracted aug-cc-pVQZ basis set is used for H, F, and Cl and uncontracted dyall.ae4z for Br, I, and At. It can be seen that the error in the non-relativistic

FNS-IP-EOM-CCSD method progressively increases on going to heavier elements. The error in non-relativistic FNO-IP-EOM-CCSD shows an error of 10.4 percent (See Figure 12) with respect to the experimental reference for HAt. The use of the four-component DC Hamiltonian leads to an improvement in the agreement of the valence IP value with the experimental results, even for the HF. The changes are even more drastic for molecules containing heavier elements. The use of the non-relativistic Hamiltonian leads to underestimation of the valence IP value for all the molecules, except for the HF, where it overestimates. The use of non-relativistic Hamiltonian gives an error of 10.4 percent with respect to the experiment for HAt. The use of scalar relativistic spin-free X2C (SF-X2C) Hamiltonian leads to an improvement over the non-relativistic one. However, the accuracy is not sufficient for heavier elements, and FNO-IP-EOM-CCSD based on SF-X2C Hamiltonian shows an error of 9.62 percent with respect to the experimental value for HAt. The error gets reduced to 0.28 percent on using a four-component DC Hamiltonian. This shows that the spin-orbit coupling interaction plays a significant role in determining the accuracy of valence IP values in the IP-EOM-CCSD method, especially for systems containing heavier elements. The FNS-IP-EOM-CCSD method also agrees well with the experiment for the splitting in the valence IP values for HCl and HBr.

Table 5 presents the comparison of results in the FNS (FNO)-CVS-IP-EOM-CCSD method based on four-component DC Hamiltonian, spin-free X2C, and non-relativistic Hamiltonian for the 1s K-edge core IP values of noble gas series. The uncontracted aug-cc-pCVTZ basis set is used for Ne and Ar, uncontracted dyall.ae3z basis set are used for Kr, Xe, and Rn. The trends in the non-relativistic results are very similar to that of the valence IP, and the non-relativistic FNO-IP-EOM-CCSD method shows a 10.75 percent error (see Figure 13) with respect to the experimental reference for Rn. However, the use of scalar relativistic SF-X2C Hamiltonian gives leads to significant improvement over non-relativistic Hamiltonian, and the CVS-FNO-IP-EOM-CCSD method based on SF-X2C Hamiltonian shows a percentage error of 0.56 for Rn. The use of a four-component DC Hamiltonian does not necessarily give better agreement with the experiment as compared to that obtained with SF-X2C Hamiltonian. The CVS-FNS-IP-EOM-CCSD method based on the four-component DC Hamiltonian shows a percentage error of 0.68 for Rn. It is well known that the spin-orbital coupling effect is very small for 1s electrons. Consequently, the use of the four-component relativistic Hamiltonian gives very similar results to that of the scalar relativistic Hamiltonian. The use of higher-order correction leads to the improvement of the DC Hamiltonian. The CVS-FNS-IP-EOM-CCSD method-based DC Hamiltonian corrected for Gaunt, Breit, and QED interactions shows a percent error of 0.08 for 1s core-ionization energy of Rn.

### 3.4 Computational Efficiency.

To demonstrate the computational efficiency of the FNS-IP-EOM-CCSD method, we have compared the timing for the FNS and canonical IP-EOM-CCSD calculation of valence IP values of HI Uncontracted aug-cc-pVTZ and dyall.ae3z basis set has been used for the hydrogen and iodine, respectively. Both of these calculations were performed serially on a dedicated workstation with two Intel(R) Xeon(R) CPU E5-2620 v4 @ 2.10GHz and 512 GB of total RAM. One valence IP root was calculated, and core electrons were included in the correlation calculation. The dimension of the virtual space in the canonical spinor basis is 462, which gets reduced to 256 in the truncated frozen natural spinor basis.

The total time taken for canonical IP-EOM-CCSD calculation for HI is two days, 23 hours, 14 minutes, and 43 seconds, which gets reduced to 11 hours, 59 minutes, and 56 seconds in the truncated frozen natural spinor basis. Figure 14 presents the time taken by the different steps in the correlation calculation in canonical and FNS-based implementation of IP-EOM-CCSD for HI It can be seen that the IP-EOM-CCSD calculation is dominated by the generation of integrals in the molecular spinor basis and the iterative solution of CCSD equations. In both steps, the use of FNS approximation leads to significant speed up over the canonical implementation. The speed up is also observed for the construction of $\bar{H}$ intermediates and Davidson's iterative diagonalization procedure. However, their contribution to the total computation time is negligible.

## 4. Conclusions:

We present the theory and implementation of a frozen natural spinor-based relativistic equation of motion coupled cluster method for ionization potential. The core-valence separation has been employed for the FNS-IP-EOM-CCSD method for core-ionization energy. The convergence of the ionization potential with respect to the size of the virtual space is much smoother in the natural spinor basis than that observed in the canonical spinor basis for both valence and core ionization energy. The natural spinors generated from the ground state MP2 wavefunction can reduce the dimension of the virtual space for the IP-EOM-CCSD method. The valence IP values converge on reaching, and core ionization potential values converge on reaching the FNS threshold of $10^{-5}$ and $10^{-6}$, respectively. The use of the frozen core approximation does not make any appreciable change in the convergence of the IP values with respect to the size of the virtual space, in contrast to that observed for ground state coupled cluster energy. The higher-order relativistic effects like Gaunt, and Breit, and QED corrections are important for the core-ionization energy. However, their contribution is negligible to the valence ionization energy.

The extension of the FNS-based approach to the excited and electron-attached state will require the use of state-specific FNS generated from a second-order approximate method for excited states. The work is in progress in that direction.

**Supplementary Material:**

Additional comparison on the convergence of valence and core ionization energy, the effect of basis set, and frozen core approximation has been provided in the supplementary material.

**Acknowledgment**

The authors acknowledge the support from the IIT Bombay, CRG and Matrix Project of DST-SERB, CSIR-India, DST-Inspire Faculty Fellowship, Prime Minister's Research Fellowship, ISRO for financial support and IIT Bombay super computational facility and C-DAC Supercomputing resources (PARAM Yuva-II, Param Bramha) for computational time.

# References

[1] I. Shavitt and R.J. Bartlett, *Many-Body Methods in Chemistry and Physics* (Cambridge University Press, 2009).

[2] D.J. ROWE, Rev Mod Phys **40**, 153 (1968).

[3] J.F. Stanton and R.J. Bartlett, J Chem Phys **98**, 7029 (1993).

[4] M. Nooijen and R.J. Bartlett, J Chem Phys **102**, 3629 (1995).

[5] J. Liu and L. Cheng, Wiley Interdiscip Rev Comput Mol Sci **11**, e1536 (2021).

[6] H. Koch and P. Jørgensen, J Chem Phys **93**, 3333 (1990).

[7] A. Landau, E. Eliav, and U. Kaldor, Advances in Quantum Chemistry **39**, 171 (2001).

[8] E. Eliav and U. Kaldor, Challenges and Advances in Computational Chemistry and Physics **11**, 113 (2010).

[9] S. Pal and D. Mukherjee, Advances in Quantum Chemistry **20**, 291 (1989).

[10] L. Meissner, J Chem Phys **108**, 9227 (1998).

[11] H. Pathak, S. Sasmal, M.K. Nayak, N. Vaval, and S. Pal, Phys. Rev. A **90**, 62501 (2014).

[12] A. Shee, T. Saue, L. Visscher, and A. Severo Pereira Gomes, J Chem Phys **149**, 174113 (2018).

[13] K.G. Dyall and K. Fægri Jr, *Introduction to Relativistic Quantum Chemistry* (Oxford University Press, 2007).

[14] J. v. Pototschnig, A. Papadopoulos, D.I. Lyakh, M. Repisky, L. Halbert, A. Severo Pereira Gomes, H.J.A. Jensen, and L. Visscher, J Chem Theory Comput **17**, 5509 (2021).

[15] Z. Cao, F. Wang, and M. Yang, J Chem Phys **145**, 154110 (2016).

[16] S. Chamoli, K. Surjuse, B. Jangid, M.K. Nayak, and A.K. Dutta, J Chem Phys **156**, 204120 (2022).

[17] X. Yuan, L. Visscher, and A.S.P. Gomes, J Chem Phys **156**, 224108 (2022).

[18] K. Hirao and H. Nakatsuji, J Comput Phys **45**, 246 (1982).

[19] P.-O. Löwdin, Phys. Rev. **97**, 1474 (1955).

[20] A.G. Taube and R.J. Bartlett, Collect Czechoslov Chem Commun **70**, 837 (2005).

[21] T.L. Barr and E.R. Davidson, Phys. Rev. A **1**, 644 (1970).

[22] A. Landau, K. Khistyaev, S. Dolgikh, and A.I. Krylov, J Chem Phys **132**, 14109 (2010).

[23] P. Pokhilko, D. Izmodenov, and A.I. Krylov, J Chem Phys **152**, 34105 (2020).

[24] L. Gyevi-Nagy, M. Kállay, and P.R. Nagy, J Chem Theory Comput **17**, 860 (2021).

[25] A. Kumar and T.D. Crawford, J Phys Chem A **121**, 708 (2017).

[26] R.A. Mata and H. Stoll, J Chem Phys **134**, 034122 (2011).

[27] D. Mester, P.R. Nagy, and M. Kallay, J Chem Phys **146**, 194102 (2017).

[28] S.D. Folkestad and H. Koch, J Chem Theory Comput **16**, 179 (2020).

[29] C. Edmiston and M. Krauss, J Chem Phys **45**, 1833 (1966).

[30] R. Ahlrichs and W. Kutzelnigg, J Chem Phys **48**, 1819 (1968).

[31] F. Neese, F. Wennmohs, and A. Hansen, J Chem Phys **130**, (2009).

[32] P.R. Nagy, G. Samu, and M. Kállay, J Chem Theory Comput **14**, 4193 (2018).

[33] M. Schwilk, D. Usvyat, and H.-J. Werner, J Chem Phys **142**, 121102 (2015).

[34] L.S. Cederbaum, W. Domcke, and J. Schirmer, Phys Rev A (Coll Park) **22**, 206 (1980).

[35] S. Coriani and H. Koch, J Chem Phys **143**, 181103 (2015).

[36] L. Halbert, M.L. Vidal, A. Shee, S. Coriani, and A.S.P. Gomes, J Chem Theory Comput **17**, 3583 (2021).

[37] A. K. Dutta, A. Manna, B. Jangid, K. Majee, K. Surjuse, S. Arora, S. Chamoli, S. Haldar, and T. Mukhopadhyay, BAGH: A quantum chemistry software package, 2021.

[38] S. Sun, T.F. Stetina, T. Zhang, H. Hu, E.F. Valeev, Q. Sun, and X. Li, J Chem Theory Comput **17**, 3388 (2021).

[39] Q. Sun, T.C. Berkelbach, N.S. Blunt, G.H. Booth, S. Guo, Z. Li, J. Liu, J.D. McClain, E.R. Sayfutyarova, S. Sharma, S. Wouters, and G.K.L. Chan, Wiley Interdiscip Rev Comput Mol Sci **8**, e1340 (2018).

[40] H.J.Aa. Jensen, R. Bast, A.S.P. Gomes, T. Saue, L. Visscher, I.A. Aucar, V. Bakken, C. Chibueze, J. Creutzberg, K.G. Dyall, S. Dubillard, U. Ekström, E. Eliav, T. Enevoldsen, E. Faßhauer, T. Fleig, O. Fossgaard, L. Halbert, E.D. Hedegård, T. Helgaker, B. Helmich-Paris, J. Henriksson, M. van Horn, M. Iliaš, Ch.R. Jacob, S. Knecht, S. Komorovský, O. Kullie, J.K. Lærdahl, C. v. Larsen, Y.S. Lee, N.H. List, H.S. Nataraj, M.K. Nayak, P. Norman, G. Olejniczak, J. Olsen, J.M.H. Olsen, A. Papadopoulos, Y.C. Park, J.K. Pedersen, M. Pernpointner, J. v. Pototschnig, R. di Remigio, M. Repiský, K. Ruud, P. Sałek, B. Schimmelpfennig, B. Senjean, A. Shee, J. Sikkema, A. Sunaga, A.J. Thorvaldsen, J. Thyssen, J. van Stralen, M.L. Vidal, S. Villaume, O. Visser, T. Winther, S. Yamamoto, and X. Yuan, (2022).

[41] D. Jana, B. Bandyopadhyay, and D. Mukherjee, Theor Chem Acc **102**, 317 (1999).

[42] K. Kozioł, I. Agustín Aucar, and G.A. Aucar, J Chem Phys **150**, 184301 (2019).

[43] M.S. Banna and D.A. Shirley, J Chem Phys **63**, 4759 (2008).

[44] A.J. Yencha, A.J. Cormack, R.J. Donovan, A. Hopkirk, and G.C. King, Chem Phys **238**, 109 (1998).

[45] M.Y. Adam, M.P. Keane, A. Naves de Brito, N. Correia, P. Baltzer, B. Wannberg, L. Karlsson, and S. Svensson, J Electron Spectros Relat Phenomena **58**, 185 (1992).

[46] A.J. Cormack, A.J. Yencha, R.J. Donovan, K.P. Lawley, A. Hopkirk, and G.C. King, Chem Phys **221**, 175 (1997).

[47] S. Rothe, A.N. Andreyev, S. Antalic, A. Borschevsky, L. Capponi, T.E. Cocolios, H. de Witte, E. Eliav, D. v Fedorov, V.N. Fedosseev, D.A. Fink, S. Fritzsche, L. Ghys, M. Huyse, N. Imai, U. Kaldor, Y. Kudryavtsev, U. Köster, J.F.W. Lane, J. Lassen, V. Liberati, K.M. Lynch, B.A. Marsh, K. Nishio, D. Pauwels, V. Pershina, L. Popescu, T.J. Procter, D. Radulov, S. Raeder, M.M. Rajabali, E. Rapisarda, R.E. Rossel, K. Sandhu, M.D. Seliverstov, A.M. Sjödin, P. van den Bergh, P. van Duppen, M. Venhart, Y. Wakabayashi, and K.D.A. Wendt, Nat Commun **4**, 1835 (2013).

[48] L. Pettersson, J. Nordgren, L. Selander, C. Nordling, K. Siegbahn, and H. Ågren, J Electron Spectros Relat Phenomena **27**, 29 (1982).

[49] J.A. Bearden and A.F. Burr, Rev Mod Phys **39**, 125 (1967).

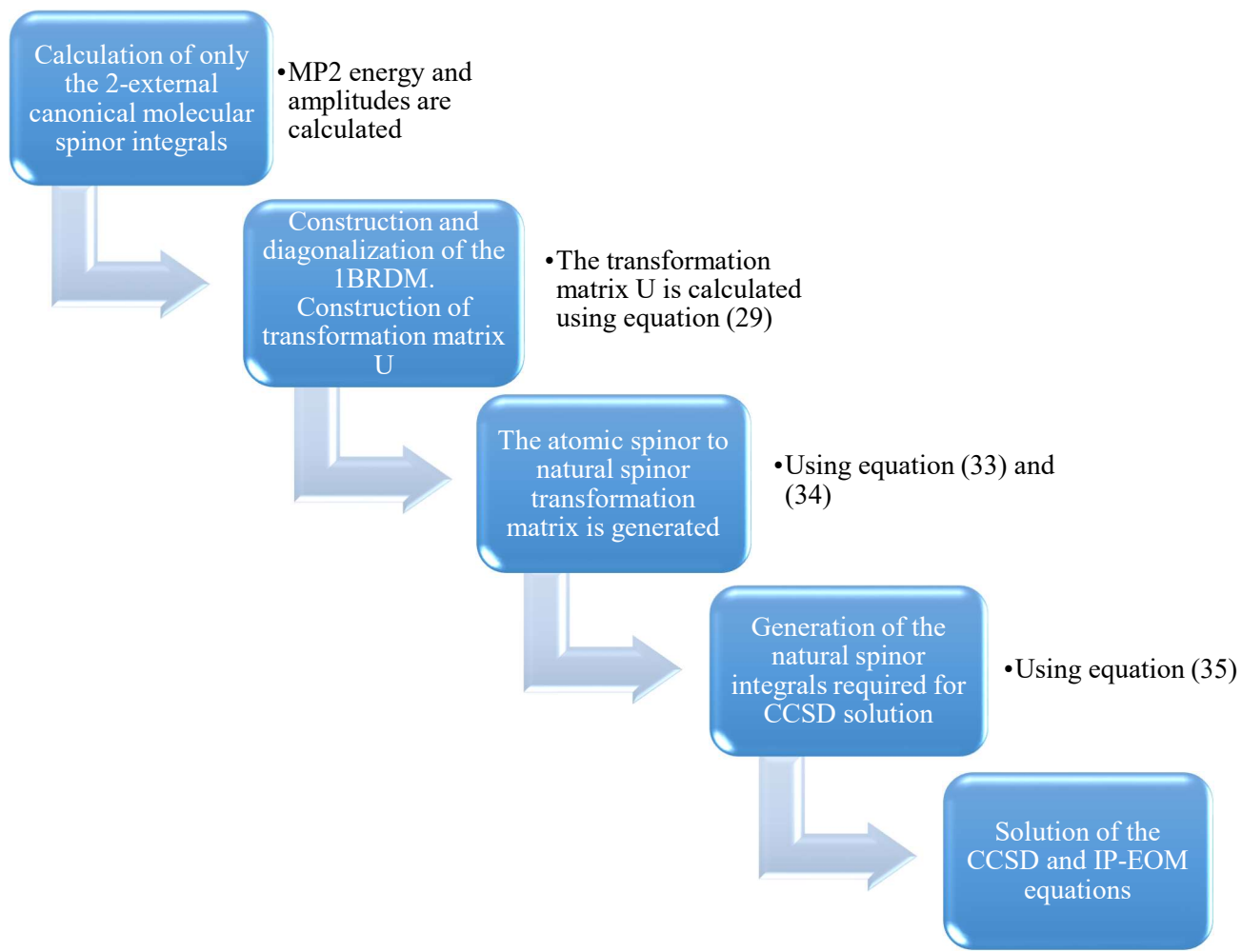

*Figure 1: A schematic description of the FNS-IP-EOM-CCSD algorithm.*

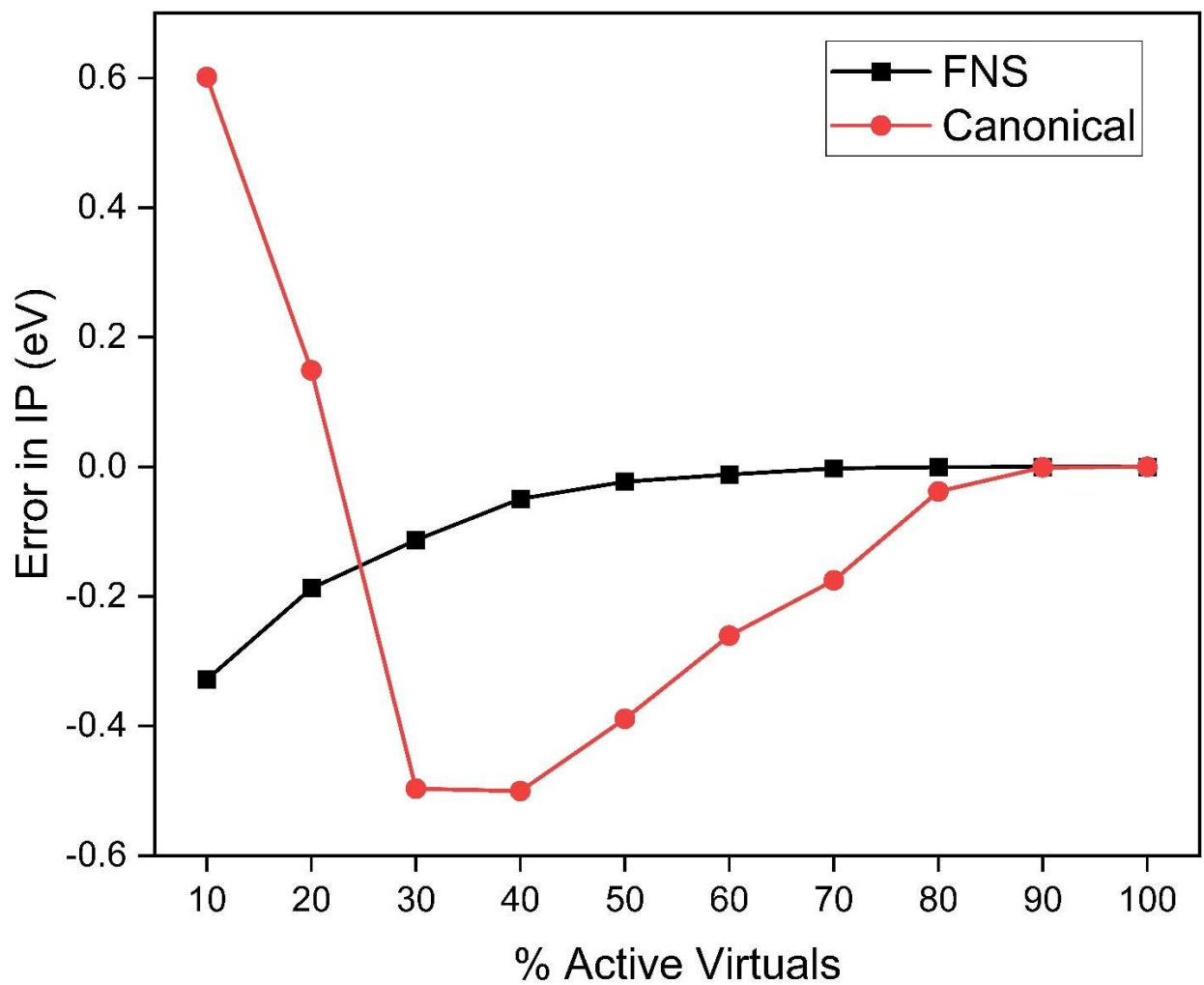

*Figure 2: Error in the first IP value of HF in IP-EOM-CCSD method with respect to the % of active virtual in canonical and FNS basis. The uncontracted aug-cc-pVTZ basis is used for both H and F.*

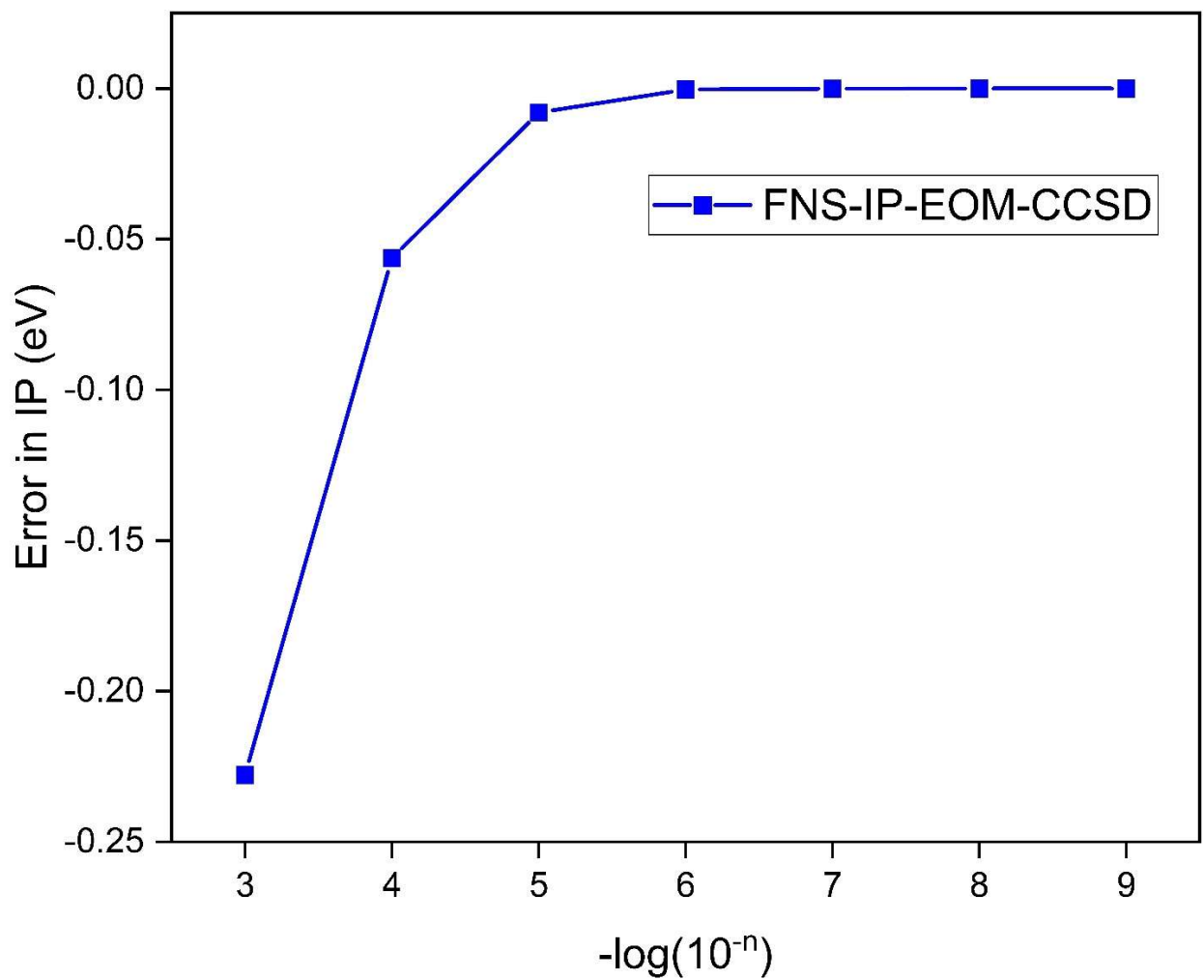

*Figure 3: The convergence of error for the first IP value of HF in the FNS-IP-EOM-CCSD method with respect to the FNS threshold. The uncontracted aug-cc-pVTZ basis is used for both H and F. The canonical four-component IP-EOM-CCSD result with 100% active virtual has been used as the reference.*

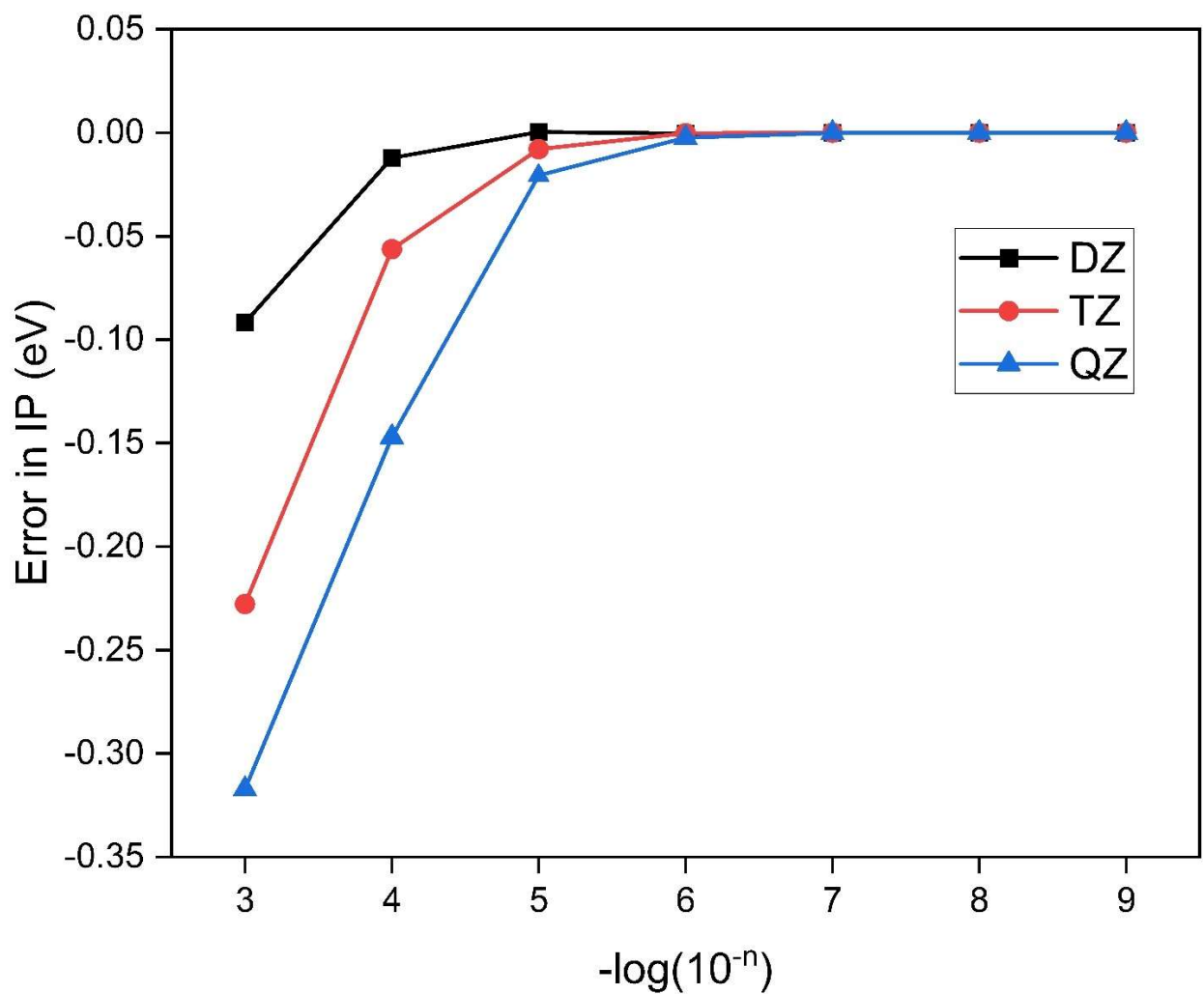

*Figure 4: The convergence of error for the first IP value of HF in the FNS-IP-EOM-CCSD method with respect to the FNS threshold with different basis sets. The uncontracted aug-cc-pVXZ (XZ = DZ, TZ, QZ) basis are used for H and F. The canonical four-component IP-EOM-CCSD results with 100% active virtual have been used as the reference.*

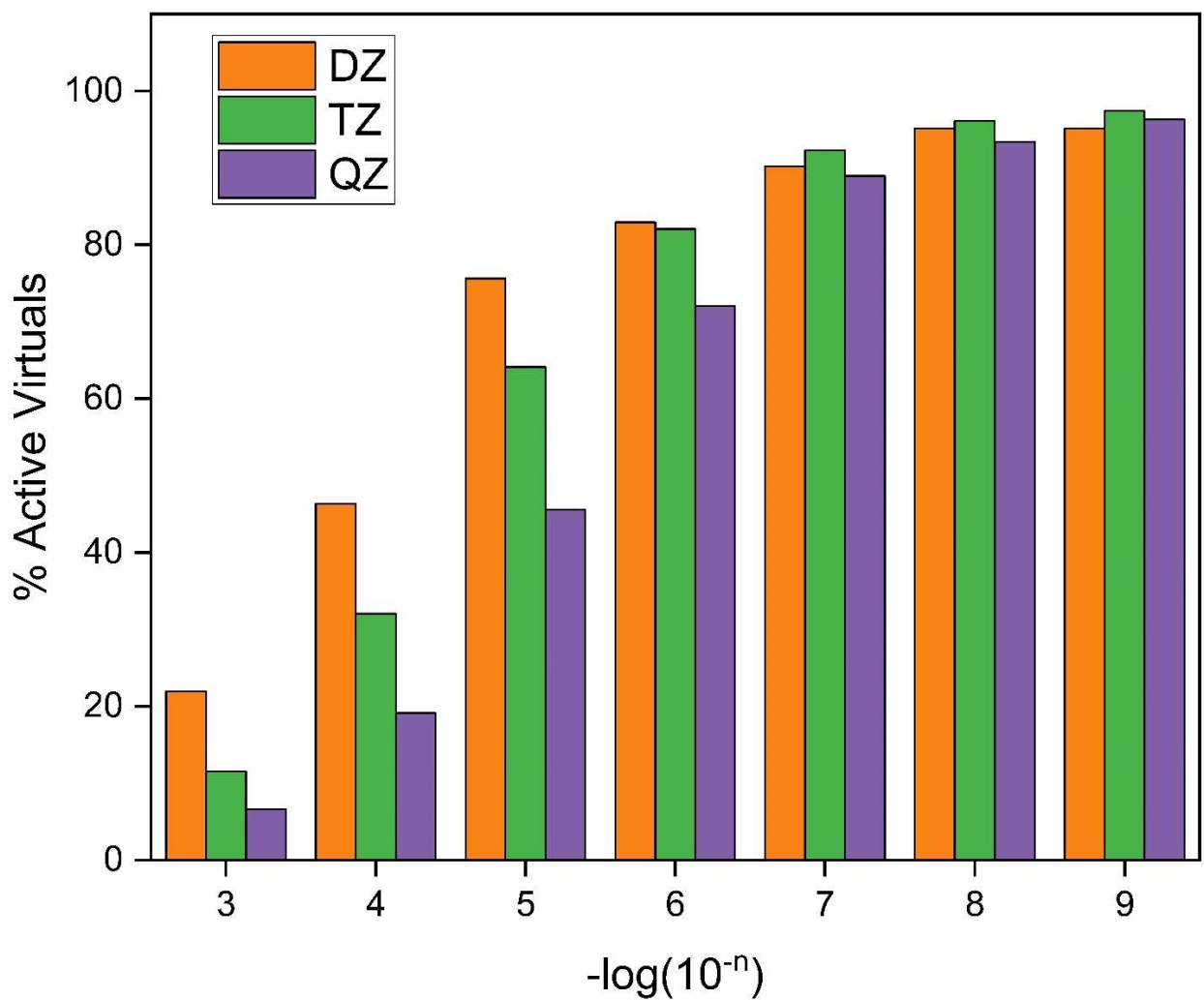

*Figure 5: The percentage of active virtuals for HF in FNS-IP-EOM-CCSD with respect to the FNS threshold with different basis sets. The uncontracted aug-cc-pVXZ (XZ = DZ, TZ, QZ) basis are used for H and F.*

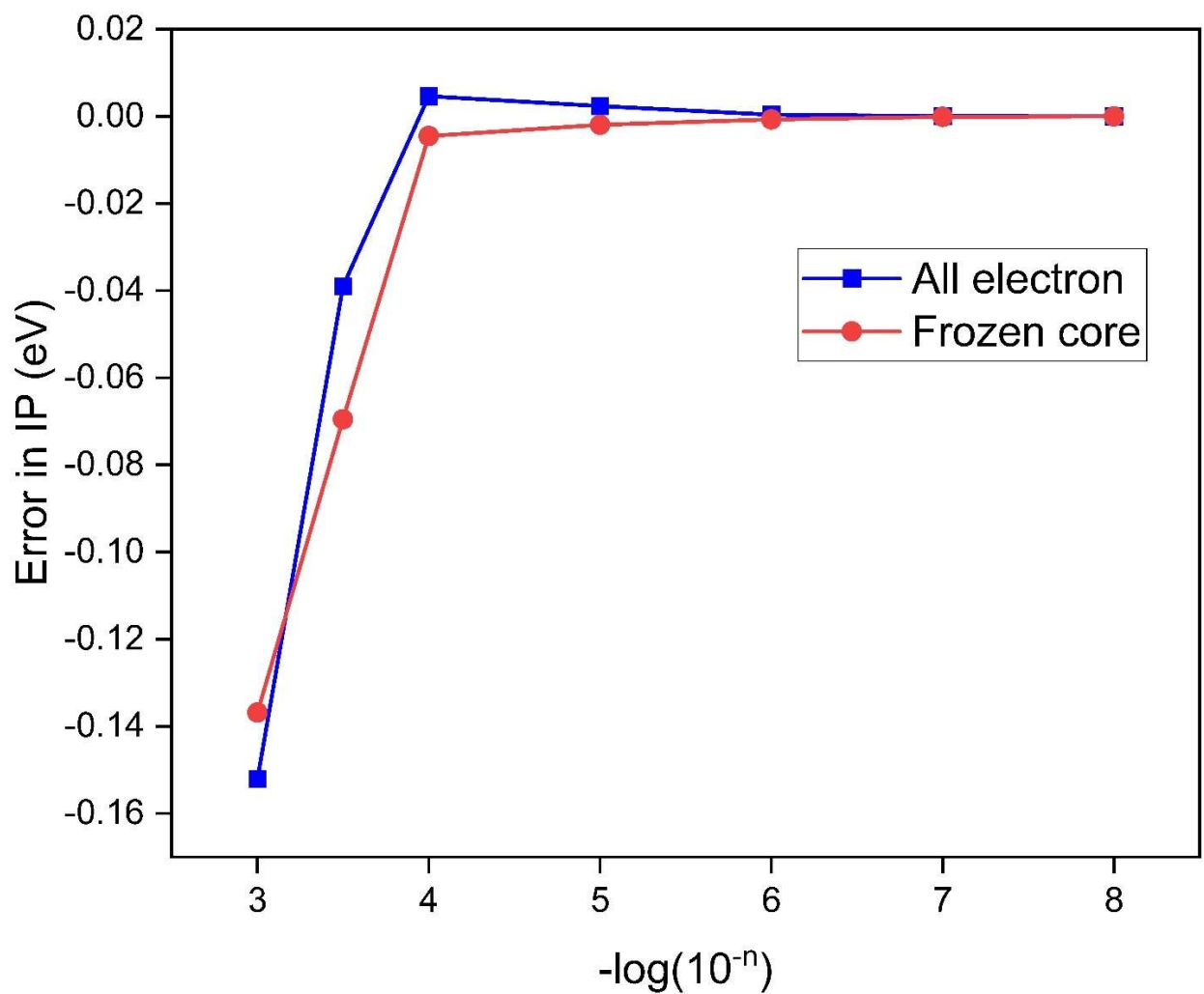

*Figure 6: Error in the FNS-IP-EOM-CCSD valence IP value with respect to the FNS threshold in all-electron and frozen core calculations for HBr. The uncontracted aug-cc-pVTZ basis is used for H and uncontracted dyall.ae3z basis is used for Br.*

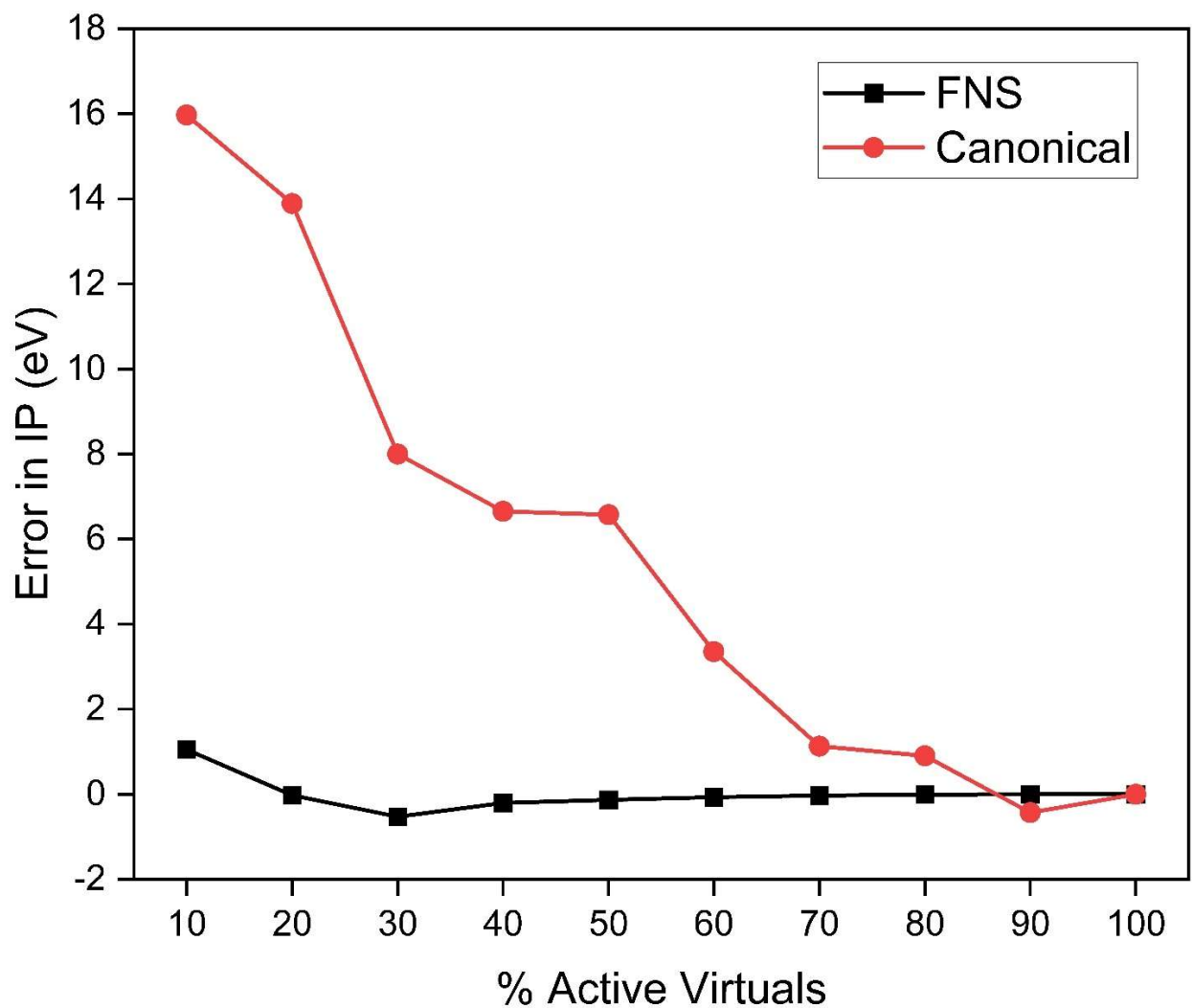

*Figure 7: Error in K-edge core IP value of HF in the CVS-IP-EOM-CCSD method with respect to the percentage of active virtuals in canonical and FNS basis. The uncontracted aug-cc-pVTZ basis is used for both H and F. The canonical four-component CVS-IP-EOM-CCSD result with 100% active virtual has been used as the reference.*

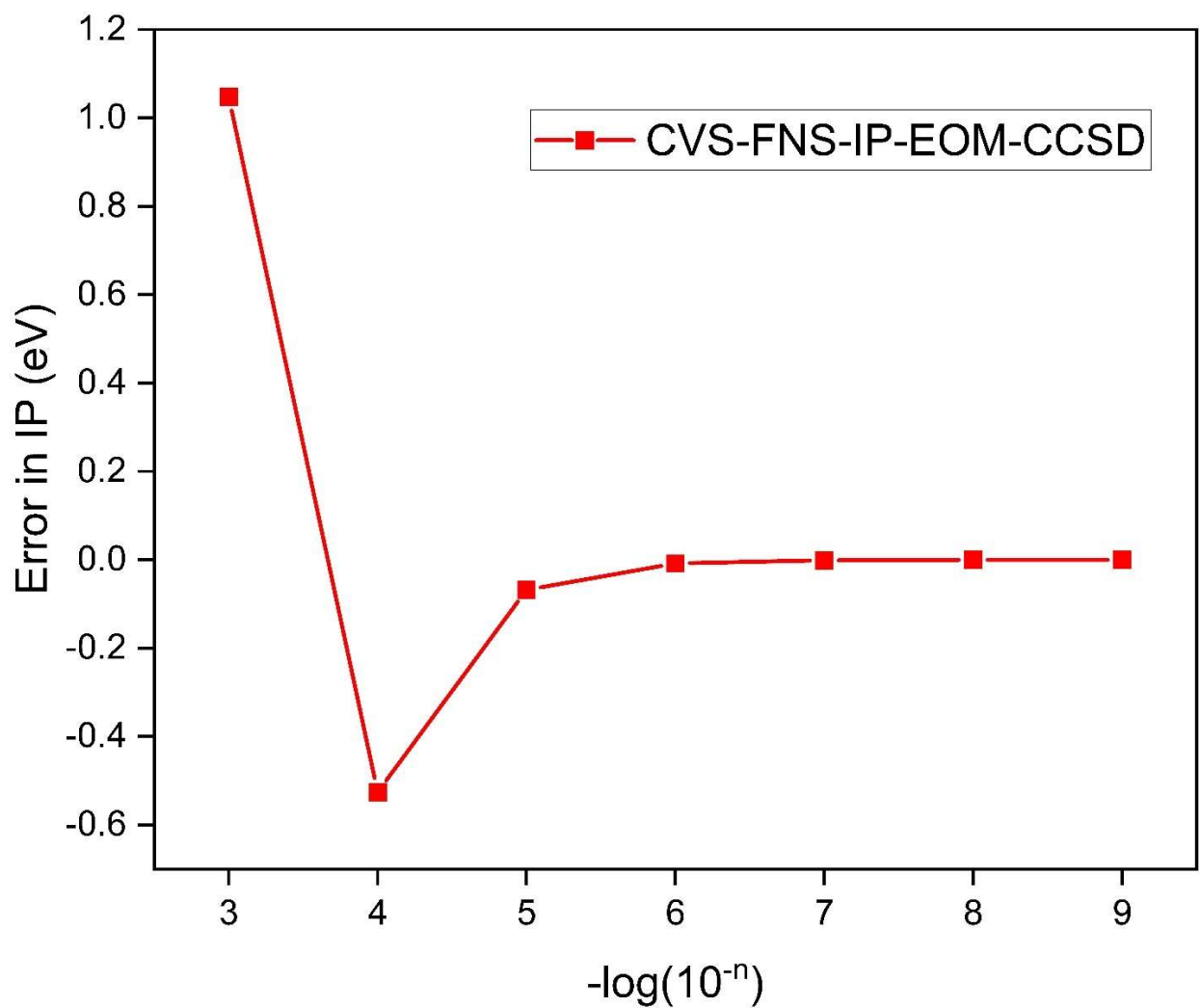

*Figure 8: The error in the CVS-FNS-IP-EOM-CCSD with respect to the FNS threshold for HF The uncontracted aug-cc-pVTZ basis is used for both H and F. The canonical four-component CVS-IP-EOM-CCSD result with 100% active virtual has been used as the reference.*

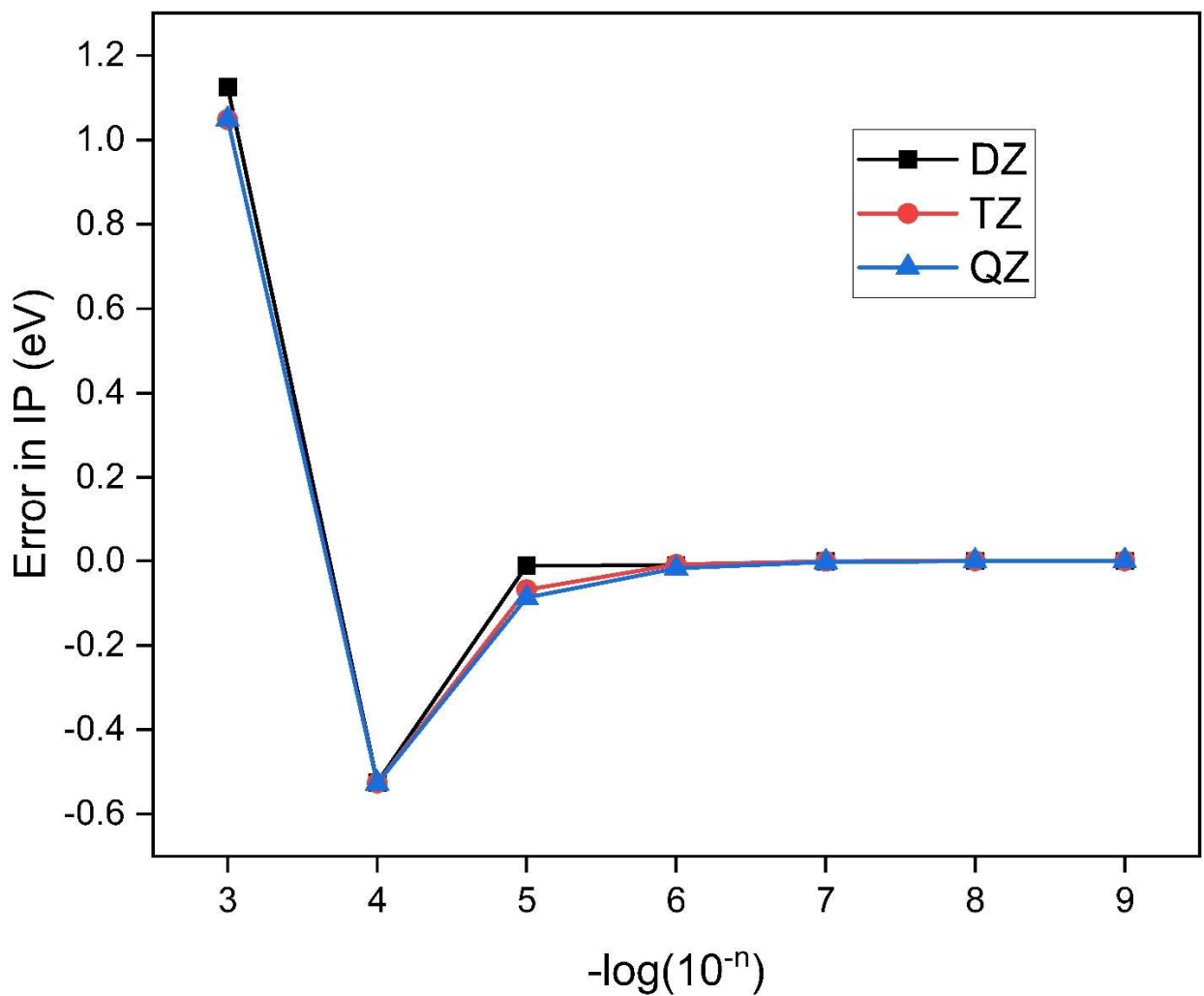

*Figure 9: The convergence of error in core IP in CVS-FNS-IP-EOM-CCSD with respect to the FNS threshold for HF in uncontracted aug-cc-pVXZ (XZ= DZ, TZ, QZ) basis set. The canonical four-component CVS- IP-EOM-CCSD results with 100% active virtual have been used as the reference.*

Table 1. FNS-IP-EOM-CCSD results for HX (X = F, Cl, Br, I, At) with different basis sets at an FNS threshold of 10$^{-5}$. The uncontracted aug-cc-pVXZ (X=D, T, and Q) basis is used for H, F, and Cl and uncontracted dyall.aexz (x=2, 3, and 4) basis is used for Br, I, and At.

| Molecule | Ionization state | DZ. | TZ. | QZ | Exp |
|---|---|---|---|---|---|
| HF  | 2 π  | 15.894 | 16.029 | 16.106 | 16.120[43] |
|     | 1 π  | 15.931 | 16.068 | 16.145 |  |
|     | 3 σ  | 19.936 | 19.984 | 20.042 | 19.890[43] |
| HCl | 4 π  | 12.505 | 12.671 | 12.768 | 12.745[44] |
|     | 3 π  | 12.587 | 12.754 | 12.852 | 12.830[44] |
|     | 5 σ  | 16.625 | 16.726 | 16.791 |  |
| HBr | 8 π  | 11.334 | 11.573 | 11.689 | 11.645[45] |
|     | 7 π  | 11.670 | 11.909 | 12.025 | 11.980[45] |
|     | 8 σ  | 15.582 | 15.728 | 15.811 | 15.650[45] |
| HI  | 12 π | 10.079 | 10.303 | 10.425 | 10.388[46] |
|     | 11 π | 10.754 | 10.968 | 11.091 | 11.047[46] |
|     | 11 σ | 14.268 | 14.380 | 14.461 |  |
| HAt | 18 π | 8.982  | 9.165  | 9.290  | 9.317[47] |
|     | 17 π | 10.763 | 10.914 | 11.035 |  |
|     | 15 σ | 14.068 | 14.145 | 14.240 |  |

Table 2. The basis set convergence of 1S core IP values in the CVS-FNS-IP-EOM-CCSD method using aug-cc-pCVXZ (X=D to Q) basis for Ne and Ar and uncontracted dyall.aexz(x=2, 3, and 4 ) basis are used for Kr and Xe. [a]

| Atoms | DZ. | TZ. | QZ. | Experimental |
|---|---|---|---|---|
| Ne | 873.26 | 872.67 | 872.57 | 870.21[48] |
| Ar | 3213.20 | 3212.93 | 3212.87 | 3202.9 ± 0.3[49] |
| Kr | 14363.36 | 14363.31 | 14363.26 | 14325.6 ± 0.8[49] |
| Xe | 34694.81 | 34694.54 | 34694.42 | 34561.4 ± 1.1[49] |

[a] Threshold of $10^{-6}$ is used.

**Table 3.** The effect of higher order relativistic correction(Gaunt and Breit) and QED correction in the CVS-FNS-IP-EOM-CCSD K-edge core-ionization energies of Noble gas atoms[a,b].

| Atoms | DC | DCG | DCB | DCB+QED[b] | Experimental |
|---|---|---|---|---|---|
| Ne | 872.67 | 872.37 | 872.37 | - | 870.21[48] |
| Ar | 3212.93 | 3210.64 | 3210.73 | - | 3202.9 ± 0.3[49] |
| Kr | 14363.39 | 14340.45 | 14341.86 | 14330.69 | 14325.6 ± 0.8[49] |
| Xe | 34692.67 | 34607.76 | 34613.86 | 34571.19 | 34561.4 ± 1.1[49] |
| Rn | 99075.93 | 98662.77 | 98692.73 | 98488.38 | 98404 ± 14.1[49] |

[a] Threshold of $10^{-6}$ is used.
[b] Basis set used Ne, Ar :unc-aug-cc-pVTZ ; Kr, Xe, Rn: dyall.ae3z

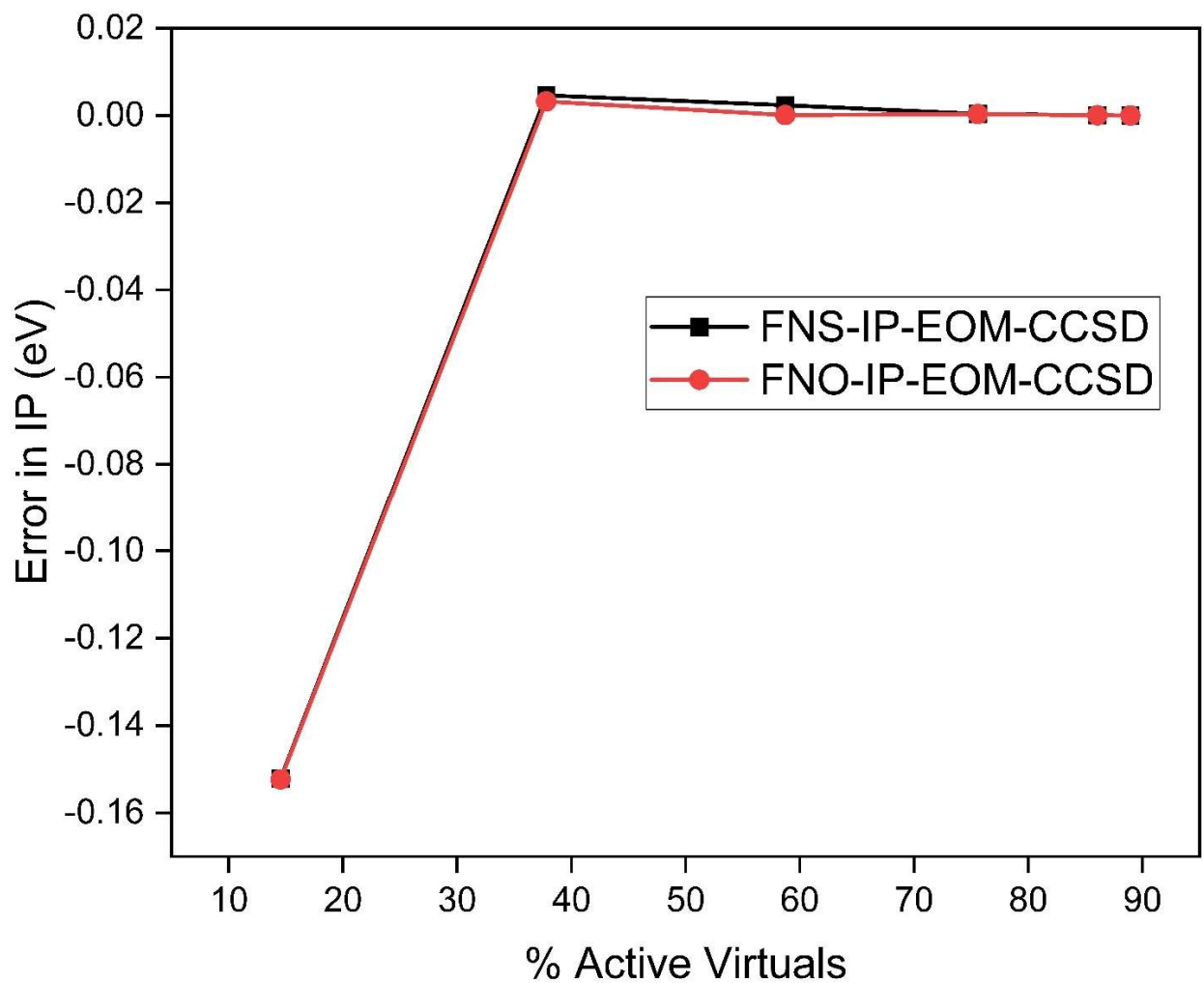

*Figure 10: Comparison of error in the valence IP values in relativistic four-component FNS-IP-EOM-CCSD and non-relativistic FNO-IP-EOM-CCSD with respect to % active virtuals for HBr. The uncontracted aug-cc-pVTZ basis set is used for H and uncontracted dyall.ae3z basis is used for Br.*

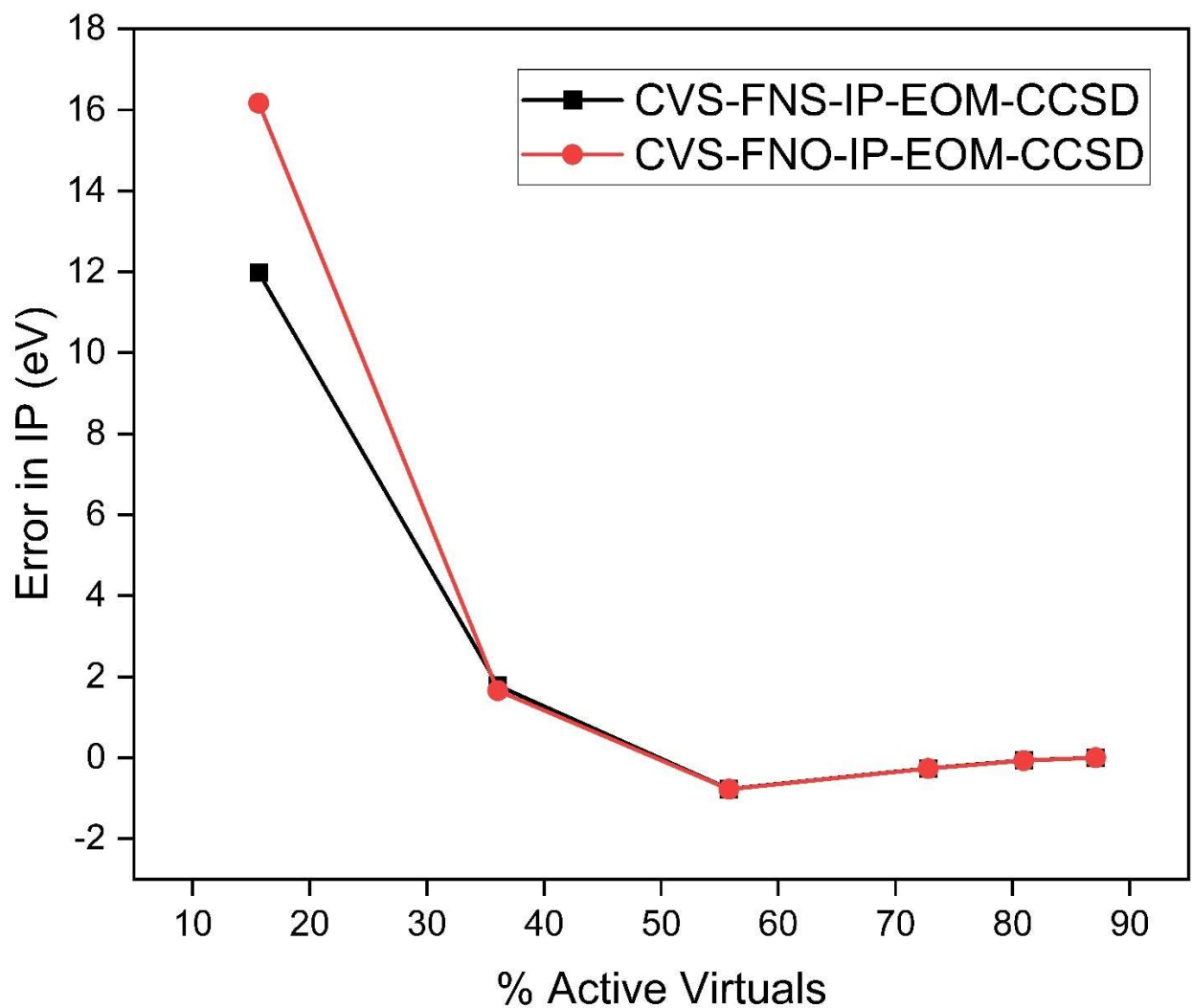

*Figure 11: Comparison of error in the 1S core IP values between relativistic four-component CVS-FNS-IP-EOM-CCSD and non-relativistic CVS-FNO-IP-EOM-CCSD with respect to % active virtuals for Kr. The uncontracted dyall.ae3z basis set is used for Kr.*

Table 4. Comparison of results in FNS(FNO)-IP-EOM-CCSD method based on four-component DC, scalar relativistic, and non-relativistic Hamiltonian for valence IP of HX (X = F, Cl, Br, I, and At)[a]

| Molecule | Ionization state | FNS-IP-EOM-CCSD | FNO-IP-EOM-CCSD | FNO-X2C-IP-EOM-CCSD | Exp |
|---|---|---|---|---|---|
| HF  | 2 π  | 16.106 | 16.162 | 16.151 | 16.120[43] |
|     | 1 π  | 16.145 | 16.162 | 16.151 |  |
|     | 3 σ  | 20.042 | 20.067 | 20.063 | 19.890[43] |
| HCl | 4 π  | 12.768 | 12.841 | 12.826 | 12.745[44] |
|     | 3 π  | 12.852 | 12.841 | 12.826 | 12.830[44] |
|     | 5 σ  | 16.791 | 16.802 | 16.804 |  |
| HBr | 8 π  | 11.689 | 11.892 | 11.861 | 11.645[45] |
|     | 7 π  | 12.025 | 11.892 | 11.861 | 11.980[45] |
|     | 8 σ  | 15.811 | 15.782 | 15.802 | 15.650[45] |
| HI  | 12 π | 10.425 | 10.802 | 10.759 | 10.388[46] |
|     | 11 π | 11.091 | 10.802 | 10.759 | 11.047[46] |
|     | 11 σ | 14.461 | 14.350 | 14.423 |  |
| HAt | 18 π | 9.290  | 10.286 | 10.214 | 9.317[47] |
|     | 17 π | 11.035 | 10.286 | 10.214 |  |
|     | 15 σ | 14.240 | 13.691 | 13.841 |  |

[a]Uncontracted aug-cc-pVQZ basis set is used for H, F, and Cl and uncontracted dyall.ae4z basis for Br, I, and At.

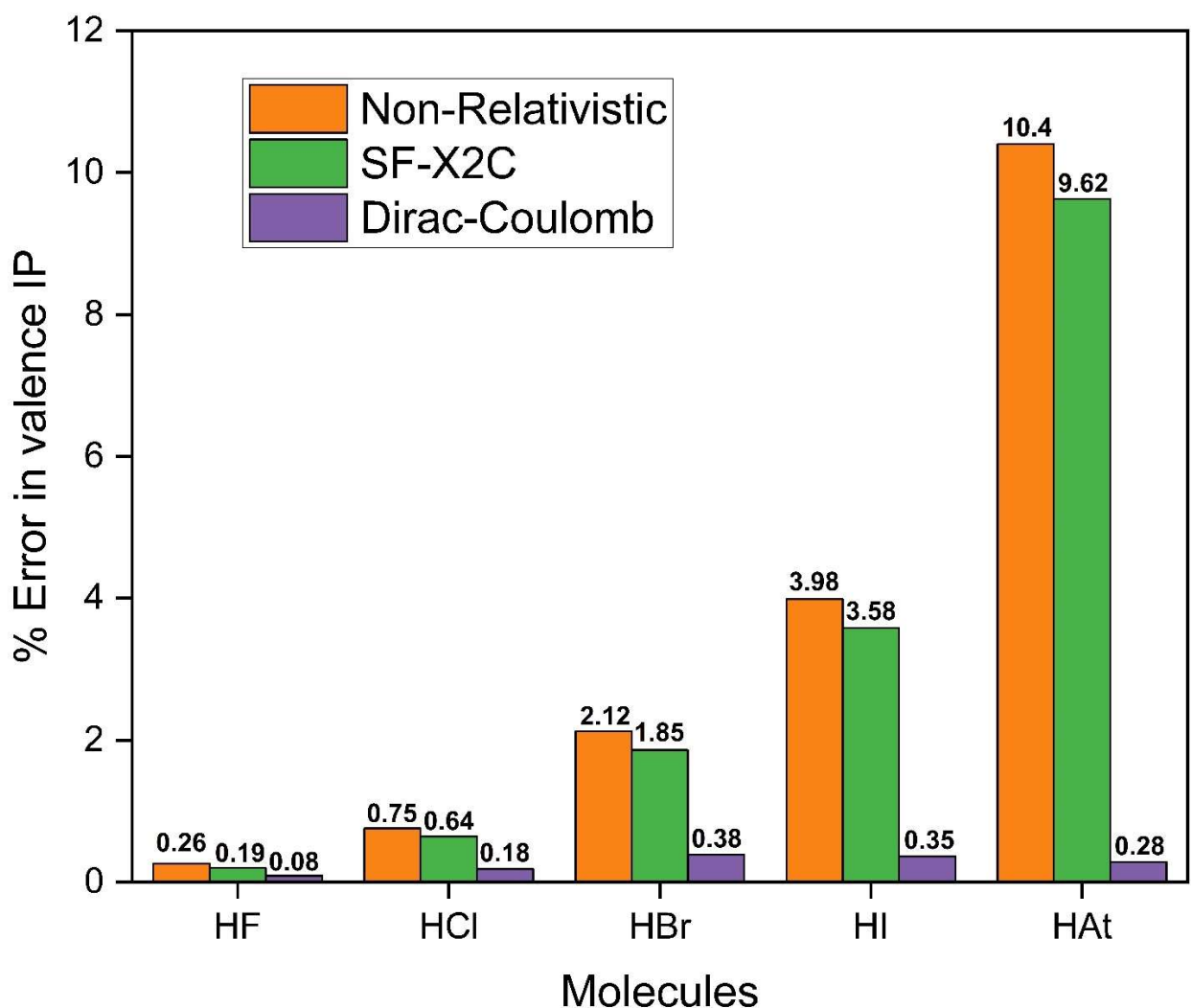

*Figure 12: The percentage error in frozen natural spinor (orbital) based implementation of IP-EOM-CCSD method with the four component DC, spin-free X2C, and non-relativistic Hamiltonian with respect to experimental results. The uncontracted aug-cc-pCVQZ basis set is used for H, F and Cl, and dyall.ae4z is the basis set used for Br, I, and At.*

Table 5. Comparison of results in FNS(FNO)-CVS-IP-EOM-CCSD method based on four-component DC, scalar relativistic, and non-relativistic Hamiltonian for 1S core IP of a noble gas atom.

| Atoms /Molecules | DC | Corrected DC | Non-Relativistic | SF-X2C | Experimental |
|---|---|---|---|---|---|
| Ne | 872.67 | 872.37 | 871.13 | 872.15 | 870.21[48] |
| Ar | 3212.93 | 3210.73 | 3198.89 | 3211.69 | 3202.9 ± 0.3[49] |
| Kr | 14363.39 | 14330.69 | 14105.45 | 14354.59 | 14325.6 ± 0.8[49] |
| Xe | 34692.67 | 34571.19 | 33256.60 | 34663.89 | 34561.4 ± 1.1[49] |
| Rn | 99075.93 | 98488.38 | 87820.05 | 98957.05 | 98404 ± 14.1[49] |

[a] Ne, Ar :unc-aug-cc-pCVTZ ; Kr, Xe, Rn: dyall.ae3z

[b] Corrected DC= DC+ Breit correction for Ne and Ar, DC+Breit+QED correction for Kr, Xe, and Rn.

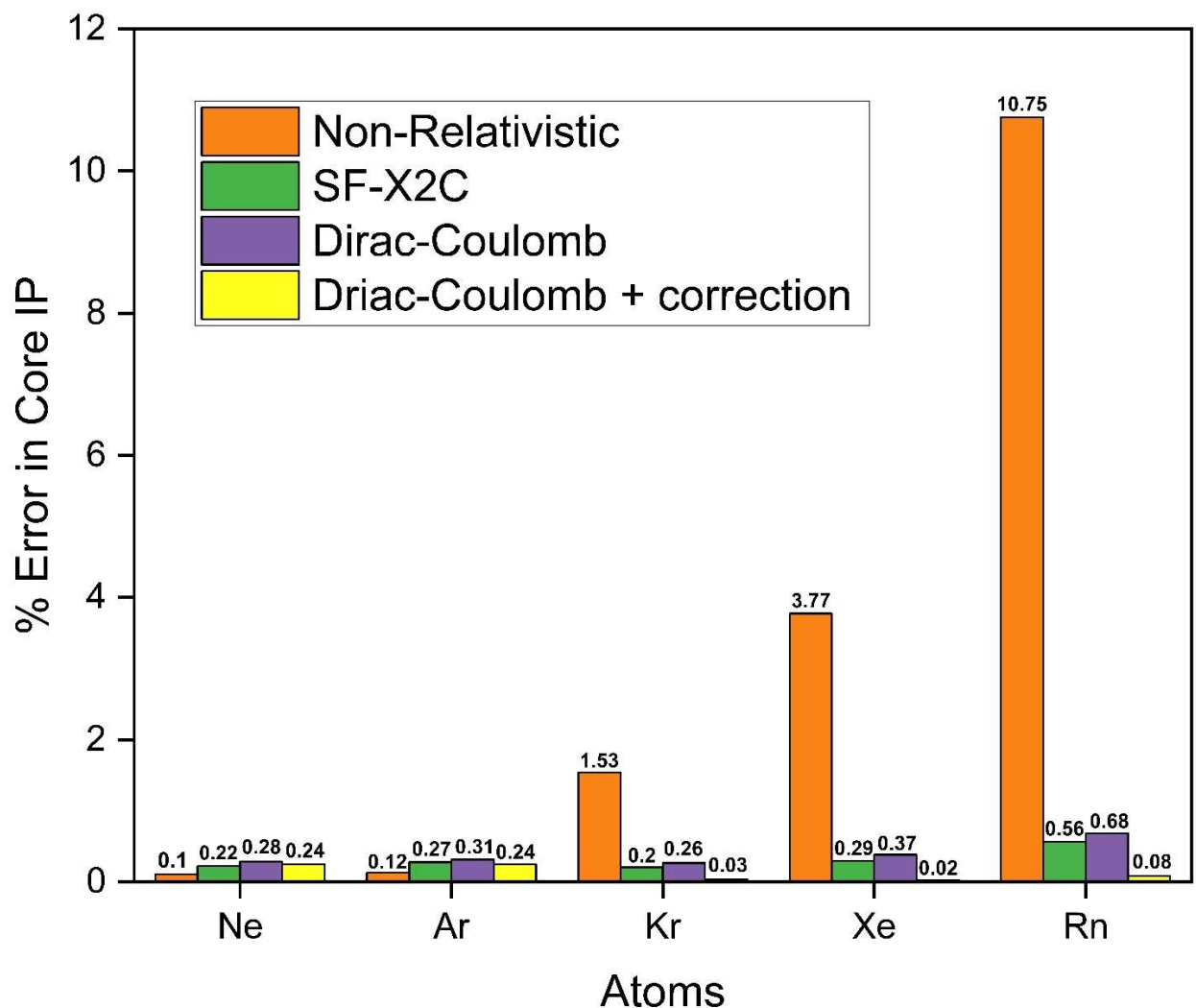

Figure 13: The percentage error in frozen natural spinor (orbital) based implementation of the CVS-IP-EOM-CCSD method with the four component DC, corrected-four component DC, spin-free X2C, and non-relativistic Hamiltonian with respect to experimental results. The uncontracted aug-cc-pCVTZ basis set is used for Ne and Ar, and dyall.ae4z basis set is used for Kr, Xe, and Rn.

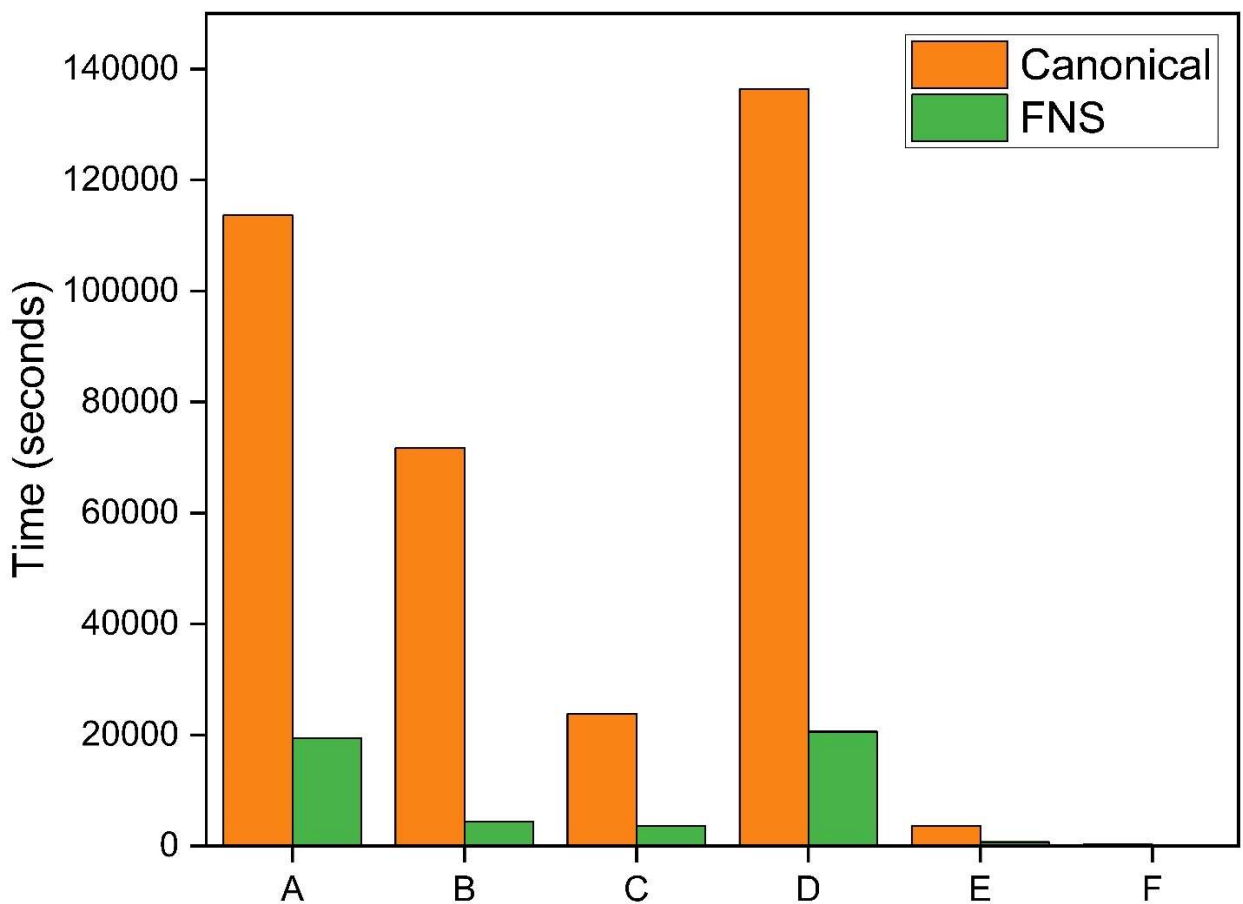

Figure 14: Comparison for time taken by the different steps in correlation calculation in canonical and FNS based implementation of IP-EOM-CCSD method for HI. The uncontracted aug-cc-pVTZ basis set is used for H and uncontracted dyall.ae3z is the basis set used for I. Processes A, B, C, D, E, and F are as follows:

A: Total time taken for generation of molecular spinor integrals
B: Time taken to generate 4 particle integrals.
C: Time taken to generate 3 particle 1 hole integrals.
D: Time taken for solution of the coupled cluster equations.
E: Time taken for $\bar{H}$ dressing.
F: Time taken for Davidson diagonalization.

Total time taken for canonical IP-EOM-CCSD calculation for HI was 2 days 23.0 hours 14.0 minutes and 42.884 seconds, i,e. 256482.884 seconds.

Total time taken for FNS-IP-EOM-CCSD calculation for HI was 11.0 hours 59.0 minutes and 55.59 seconds. i,e. 43195.59 seconds.

Both of these calculations were performed serially on a dedicated workstation using 10000 MB/core memory. The workstation has two Intel(R) Xeon(R) CPU E5-2620 v4 @ 2.10GHz and 512 GB RAM in total.